\documentclass[final,oneside,openany]{book}
\usepackage{chapterbib}
\usepackage{comment}
\usepackage{textcomp}
\usepackage{amsmath,amssymb}
\usepackage{graphicx}
\usepackage{booktabs}
\usepackage{longtable}
\usepackage{comment}
\usepackage{array}
\usepackage{tabularx}
\usepackage{multirow}
\usepackage{textgreek}
\usepackage{enumitem}
\usepackage{hyperref}
\usepackage{geometry}
\usepackage{caption}
\usepackage{url}
\usepackage{tcolorbox}
\usepackage{etoolbox}
\usepackage[T1]{fontenc}
\usepackage{tikz}
\usetikzlibrary{arrows}
\usepackage{float}
\usepackage{xcolor}
\usepackage[numbers,sectionbib]{natbib}
\newcommand{\chapterauthor}[1]{%
  \gdef\currentchapterauthor{#1}
}
\newcommand{\currentchapterauthor}{}
\author{Editors: Hema A. Murthy, Shrikanth Narayanan, and Mriganka Sur, Co-editors: R Gowriprasad, Kadiri Sudarsana Reddy}
\title{Electro Encephalogram Signals and Cognition}

\usepackage{titlesec}
\titleformat{\chapter}[display]
  {\normalfont\huge\bfseries}
  {}{0pt}{\Huge}
\titlespacing*{\chapter}{0pt}{0pt}{24pt}

\begin{document}
\frontmatter

\newpage


\begingroup
\renewcommand{\contentsname}{Contents}
\endgroup

\mainmatter

\makeatletter
\providecommand{\@makespecialcolbox}{%
  \setbox\@outputbox\vbox{\unvbox\@outputbox}%
}
\makeatother

\newtcolorbox{bestpractice}{
  colback=blue!5!white,
  colframe=blue!40!black,
  fonttitle=\bfseries,
  title={Best Practice Tip},
  boxrule=0.5pt,
  arc=2mm,
  left=6pt, right=6pt, top=6pt, bottom=6pt
}
\chapterauthor{Parsa Razmara, Woojae Jeong, Aditya Kommineni, Raymundo Cassani, Richard Leahy, and Takfarinas Medani}
\chapter{Best Practices in EEG Analysis: Preprocessing, Modeling, and Machine Learning}
\addtocontents{toc}{\protect\hspace*{1.5em}\textit{\currentchapterauthor}\par}
\label{ch:eeg-best-practices}

\noindent\textbf{Parsa Razmara\textsuperscript{1}, Woojae Jeong\textsuperscript{1}, Aditya Kommineni\textsuperscript{1}, Raymundo Cassani\textsuperscript{2},
Richard Leahy\textsuperscript{1},
and Takfarinas Medani\textsuperscript{1}}

\medskip
\noindent\textsuperscript{1}University of Southern California \quad \textsuperscript{2}McGill University

\section{Introduction and Scope}
\label{sec:introduction}

Electroencephalography (EEG) provides millisecond-scale access to human brain dynamics but presents substantial methodological challenges arising from low signal-to-noise ratio, volume conduction, and heterogeneous recording conditions. This chapter presents a comprehensive and practice-oriented overview of best practices in EEG preprocessing, computational modeling, statistical inference, and machine learning.

Section~\ref{sec:preprocessing} addresses principled preprocessing strategies, including identification and mitigation of physiological and non-physiological artifacts, filtering design and parameter selection, bad-channel detection and interpolation, referencing schemes, and artifact correction using ICA, subspace-based methods, and model-informed approaches. Special considerations for simultaneous EEG-fMRI acquisition are discussed, including gradient and pulse artifact correction. The section concludes with guidelines for standardized data organization using BIDS-EEG and structured derivative management to ensure reproducibility and transparency.

Section~\ref{sec:computational-modeling} introduces computational modeling approaches for extracting meaningful neural structure from mixed scalp recordings. Classical event-related potential analysis, time--frequency methods, connectivity metrics, source localization, and multivariate decoding are reviewed. Emphasis is placed on methodological assumptions, interpretability, and appropriate statistical inference, including permutation testing and multiple-comparison correction in high-dimensional EEG data.

Section~\ref{sec:machine-learning} examines machine learning approaches for EEG, ranging from feature-based classical classifiers to deep learning architectures and emerging foundation models. The discussion highlights challenges specific to EEG, including inter-subject variability, limited dataset sizes, and evaluation strategies for within-subject and cross-subject generalization. Model evaluation metrics and benchmarking practices are addressed, with attention to avoiding data leakage and ensuring fair comparison across methods.

Throughout the chapter, reproducibility is treated as a foundational principle. Transparent reporting of preprocessing parameters, preservation of raw data, standardized derivative storage, and adherence to FAIR data principles are emphasized as prerequisites for robust statistical inference and reliable machine learning applications in EEG research.

This chapter is intended as a self-contained reference for neuroscientists and neuroengineers, integrating methodological rigor with practical guidance to support reproducible and robust EEG research.

\medskip
\noindent\textbf{Keywords:} EEG preprocessing; artifacts; filtering; ICA; BIDS-EEG; time--frequency analysis; statistical testing; EEG machine learning; feature extraction; deep learning; reproducibility

\section{Preprocessing EEG Signals}
\label{sec:preprocessing}

Preprocessing is a critical first stage in EEG data analysis, aimed at enhancing signal quality by removing or correcting various artifacts and noise. EEG recordings are notoriously prone to contamination because the signals of interest (microvolt-level scalp potentials from neuronal activity) are easily obscured by other biological and environmental potentials. This section details best practices for cleaning EEG data: identifying major noise sources, applying appropriate filters, handling bad electrodes and referencing, removing artifacts using methods such as ICA and regression, segmenting data into epochs with baseline correction, and implementing objective trial rejection criteria. We also emphasize standardized data organization (using the Brain Imaging Data Structure for EEG, or BIDS-EEG) and reproducibility, including pipeline transparency and quality control. As discussed in the Software Pipelines chapter, efficient preprocessing workflows can be automated and standardized; here, we focus on the methodological choices that ensure valid and reliable EEG analyses.

\subsection{Common EEG Noise Sources and Artifacts}
\label{sec:noise-sources}

Real EEG data contain numerous artifacts---signals that do not originate from brain activity---which must be managed to avoid misleading results~\cite{bitbrain2025}. These artifacts can be broadly categorized into \textbf{physiological artifacts} (generated by the subject's body) and \textbf{non-physiological artifacts} (stemming from the recording hardware or environment). Here, we survey the most common artifact types:

\begin{itemize}[leftmargin=*]
  \item \textbf{Eye Movements and Blinks (EOG artifacts):} The human eye acts as an electrical dipole. Blinks and vertical eye movements produce large voltage deflections, especially at frontal scalp sites (often 100--200~$\mu$V). In the frequency domain, blink artifacts concentrate power in the delta ($<$4~Hz) and theta (4--7~Hz) bands, potentially overlapping with the low-frequency neural signals. Lateral eye movements (saccades) create characteristic biphasic ``sawtooth'' patterns, prominent at temporal electrodes and affecting frequencies up to $\sim$20~Hz~\cite{bitbrain2025,brainproducts2022}. Eye artifacts can mimic or obscure neural activity (for example, consistent blinks after a stimulus could be mistaken for evoked responses).

  \item \textbf{Muscle Activity (EMG artifacts):} Facial, neck, and scalp muscle contractions generate broadband high-frequency noise (20--300~Hz and beyond) that can contaminate the EEG. Bruxism (jaw clenching), frowning, or even subtle facial tension can introduce bursts of electromyographic activity observable across many channels. EMG artifacts often manifest as increased power in the beta ($>$20~Hz) and gamma bands, but strong muscle artifacts (such as teeth clenching) can overwhelm even lower frequencies and propagate widely over the scalp. Sustained neck and shoulder tension produces low-frequency drifts and high-frequency noise, especially near posterior leads and reference electrodes (e.g., mastoids).

  \item \textbf{Cardiac and Pulse Artifact:} The heartbeat can induce slow periodic artifacts via two mechanisms: (1) the electrocardiographic (ECG) signal itself, if picked up by electrodes (especially those near major arteries or if an ECG lead is recorded), and (2) slight head movements or impedance changes due to blood pulsation, sometimes called the ballistocardiogram in EEG-fMRI contexts. Pulse artifacts are usually small ($\sim$tens of microvolts) but can appear as rhythmic slow waves that might be mistaken for neuronal oscillations or pathological activity. They often appear in channels near the neck or face (including the mastoids), and, similar to EMG, re-referencing to an affected channel can broadcast the artifact globally. Pulse artifacts are especially salient in simultaneous EEG-fMRI recordings or in participants with hypertension.

  \item \textbf{Respiration and Sweat (Skin Potentials):} Slow fluctuations can arise from perspiration and breathing. Changes in skin conductance (due to sweating) create slowly varying potentials (0--0.5~Hz), often called sweat artifacts~\cite{brainlatam2026}. These appear as drifting baselines or transient slow waves that are not of neural origin. They tend to build up over time, especially in warm recording environments or during physical exertion or stress~\cite{brainproducts2022}. Sweat artifacts can severely distort low-frequency EEG and cause spurious DC shifts. Respiration can also modulate the EEG subtly via movement or pH changes, contributing to slow oscillatory baselines. Together, these slow artifacts manifest as baseline drifts that can bias measures of event-related potentials (ERPs) or slow-wave neural activity.

  \item \textbf{Line Noise and Electromagnetic Interference:} Mains electricity (50 or 60~Hz, depending on region) induces a pervasive artifact in EEG, often visible as a narrowband peak in power spectra. This power-line noise can couple into EEG electrodes via the environment or mains-powered equipment. Its harmonics (e.g., 100, 120,~\ldots~Hz) may also appear with lower power in the spectra. Unlike physiological artifacts, line noise is non-biological and manifests as a roughly sinusoidal interference at a fixed frequency. Without proper handling, it reduces signal-to-noise ratio (SNR), especially for analyses in the beta/gamma range. Environmental electromagnetic interference (e.g., from nearby monitors, power supplies) can similarly introduce noise.

  \item \textbf{Electrode Pops and Cable Movements:} Sudden changes in electrode impedance (for instance, if an electrode momentarily loses contact or if a cable is jostled) produce sharp transient artifacts. These electrode pops often appear as large spikes or step-like shifts that can saturate amplifier channels. Cable movements (especially in mobile EEG setups) induce motion artifacts that are typically broadband and non-stereotyped. Unlike physiological artifacts, these technical artifacts often indicate a loss of a valid signal (the data during the artifact cannot be trusted and should be discarded from the analysis). A well-tested equipment setup is recommended to mitigate such artifacts, but offline analysis and detection are needed when these artefacts occur.

  \item \textbf{Head Movement and Motion Artifacts:} Gross movements (shifting position, sudden head turns) can produce large artifacts via multiple mechanisms, such as electrode lead movement, electrode impedance changes and muscle activation. These artifacts are typically large in amplitude and broadband. In high-motion scenarios (e.g., walking with a mobile EEG), motion artifacts become a dominant noise source that is challenging to fully remove~\cite{blum2019}. Careful experimental design (instructing subjects to minimize sudden movements, or using motion sensors) can help, but some studies (e.g., mobile brain/body imaging) must accept higher residual noise.
\end{itemize}

The combined effect of these artifacts is that raw EEG often has a very low effective SNR. For example, blink or muscle artifacts can have amplitudes an order of magnitude larger than typical EEG signals (50--100~$\mu$V vs 5--20~$\mu$V). If left unaddressed, artifacts can lead to false findings (e.g., interpreting an eye movement as a frontal slow wave or muscle noise as gamma-band brain activity). Therefore, rigorous artifact handling is essential.


\begin{bestpractice}
Begin any EEG analysis by visually inspecting raw data to identify which artifacts are present and their characteristics. Plotting continuous data and noting occurrences of blinks, movement, line noise, and other artifacts will guide the choice of preprocessing steps. Additionally, collect auxiliary channels during acquisition whenever possible, for instance, EOG channels for eye movements and ECG channels for heartbeats, as these can greatly assist in artifact identification and removal (e.g., regression or targeted ICA removal of EOG/ECG-related components).
\end{bestpractice}

\begin{figure}[htbp]
  \centering
  \includegraphics[width=0.85\textwidth]{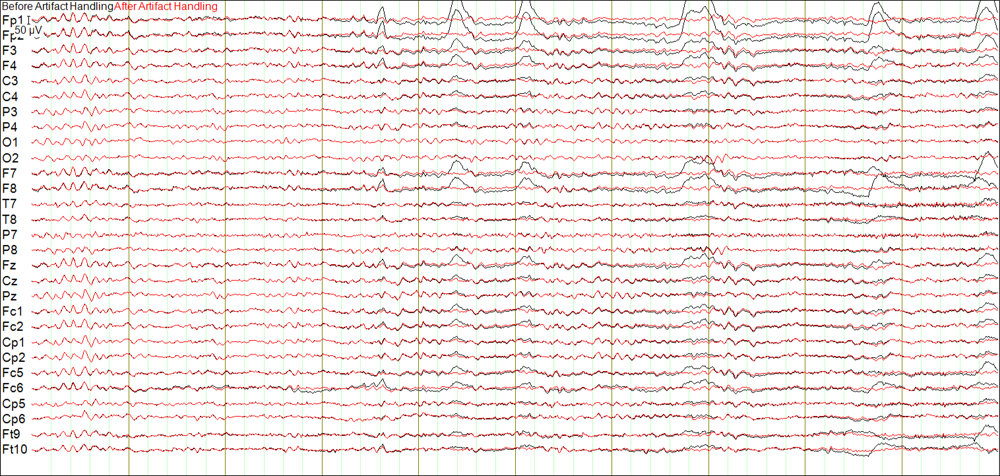}
  \caption{Raw EEG record with artifacts (black trace) and the same data after artifact correction (red trace), demonstrating the improvement in signal quality achievable with proper preprocessing.}
  \label{fig:artifact-correction}
\end{figure}
\subsection{Filtering: High-Pass, Low-Pass, Notch, and Anti-Alias Considerations}
\label{sec:filtering}

Filtering is one of the first preprocessing steps applied to EEG data. The goal is to attenuate frequency components that are outside the range of interest or dominated by noise, while preserving as much of the neural signal as possible. Filters must be used with care to avoid distorting the data (e.g., introducing phase shifts or ringing artifacts). Here, we outline best practices for filtering EEG, including cutoff frequency choices, filter types (FIR vs IIR), and special topics such as anti-aliasing.

\begin{itemize}[leftmargin=*]
  \item \textbf{High-Pass Filtering (HPF):} High-pass filters remove very slow drifts and DC offsets, including those caused by sweat and respiratory artifacts. A common practice is to high-pass the EEG at $\sim$0.1--1~Hz. However, the choice of high-pass cutoff has critical implications. Research has shown that using a cutoff frequency that is too high can distort low-frequency ERP components, potentially creating artifactual effects~\cite{mnefiltering,acunzo2012}. For example, high-pass filtering at 0.1~Hz or above can shift the time course of slow cognitive ERPs (such as the P300 or slow waves) and even generate spurious early peaks. Widmann et al.\ (2015)~\cite{widmann2015} and others recommend a very low cutoff (around 0.1~Hz) for most ERP analyses, since this preserves essentially all slow neural trends while removing only near-DC drifts. Empirical simulations have demonstrated that a 0.1~Hz high-pass filter causes minimal distortion, whereas a 0.3--0.5~Hz high-pass filter can induce noticeable latency shifts in components such as the N400 or P600 language ERPs~\cite{mnefiltering,acunzo2012}.
\end{itemize}

\begin{bestpractice}
Use the lowest high-pass frequency that is practical for your data, often 0.1~Hz for ERP studies. If slower shifts are an issue (e.g., long recordings or significant sweat artifacts), one might choose $\sim$0.5~Hz, but be cautious and report this choice. Always inspect the filter's effect on a test dataset (to ensure known slow components are not altered). If in doubt, leaving data un-high-passed and using baseline correction (see Section~\ref{sec:eegfmri}) is an option, though baseline correction and high-pass filtering overlap in their effects.
\end{bestpractice}

\begin{itemize}[leftmargin=*]
  \item \textbf{Low-Pass Filtering (LPF):} Low-pass filters remove high-frequency noise, including muscle artifacts and high-frequency line noise harmonics. A typical low-pass cutoff in cognitive EEG might be 30 or 40~Hz, assuming that signals above that (high gamma) are either not of interest or have low SNR on scalp EEG. In sleep EEG or cognitive tasks, researchers might low-pass filter at 20~Hz to focus on the delta--beta bands. When choosing a low-pass cutoff, consider the fastest neural events you wish to retain (e.g., sharp spike-wave discharges in epilepsy might require $>$70~Hz bandwidth; otherwise, a 30~Hz cutoff might smear them). Modern amplifiers usually record up to 1000~Hz, but most EEG spectral power is $<$100~Hz, except for artifacts.
\end{itemize}

\begin{bestpractice}
For typical ERP or oscillation analyses, a 30--40~Hz low-pass is common. If analyzing gamma oscillations or very high-frequency bursts, use a higher cutoff (e.g., 80~Hz) or no low-pass filter at all, and instead address muscle artifact through other means. Low-pass filters are typically linear-phase FIR filters applied in a zero-phase manner (\texttt{filtfilt} or similar) to avoid phase shifts in the data~\cite{mnefiltering}.
\end{bestpractice}

\begin{itemize}[leftmargin=*]
  \item \textbf{Notch Filtering vs.\ Alternative Line Noise Removal:} To specifically target line noise (50/60~Hz), a notch filter can be used. A notch is a very narrow band-stop filter (often 2~Hz width or so) centered on the line frequency (and sometimes its first harmonic, e.g., 120~Hz). Notch filters effectively remove the offending frequency but can introduce transient ringing and may also remove genuine neural power at that frequency.

  An alternative is to use multi-taper spectrum interpolation or regression methods to remove line noise without a classical notch. For example, the \emph{CleanLine} algorithm in EEGLAB uses a multi-taper approach to characterize and subtract sinusoidal noise~\cite{cleanline,mitra2008}, based on methods by Mitra \& Bokil (2007). Recent methods, such as spectrum interpolation, literally interpolate over the narrow spectral peak of line noise to mitigate it with minimal distortion~\cite{leske2019,dechevigne2019}. If the EEG analysis involves power spectra (especially in high beta/gamma range), one should carefully handle line noise---a notch can create a `dip' in the spectrum, and slight ripples due to the filter kernel.
\end{itemize}

\begin{bestpractice}
If line noise is strong, apply a targeted removal. A well-designed notch (e.g., 4th order Butterworth band-stop at 50/60~Hz) is acceptable, but mention its use. Alternatively, use algorithms like CleanLine, which adaptively remove periodic noise. Avoid over-aggressive notch filtering that removes wide frequency bands. Note that if data are short (e.g., a few seconds) or heavily time-locked, better to tolerate some line noise rather than distort temporal aspects.
\end{bestpractice}

\begin{itemize}[leftmargin=*]
  \item \textbf{Anti-Alias Filtering and Resampling:} Downsampling the EEG data (for instance, from 1000~Hz to 250~Hz for easier storage/processing), one must first apply a low-pass filter at the new Nyquist frequency (half the target sampling rate). This is called anti-alias filtering. For downsampling to 250~Hz, a low-pass filter at $\sim$125~Hz (often a bit lower, like 100~Hz to allow for a transition band) is used before decimation to prevent higher-frequency signals from aliasing into lower frequencies. Many EEG software packages' resampling functions do this automatically (e.g., Brainstorm, FieldTrip, MNE-Python)~\cite{tadel2011brainstorm, oostenveld2011_, gramfort2013_}. Ensure that the anti-alias filter is linear-phase to avoid distortion.
\end{itemize}

\begin{bestpractice}
Downsample only after removing high-frequency noise. If using hardware that already applies an anti-alias filter (many amplifiers have built-in analog low-pass around 250--500~Hz), you can downsample modestly (e.g., 1000 to 500~Hz) without a second filter, but for large downsampling (1000 to 250 or 128~Hz), apply an explicit anti-alias filter. Always verify that the resampled data's spectra look as expected.
\end{bestpractice}

\begin{itemize}[leftmargin=*]
  \item \textbf{Filter Design and Order:} Whenever possible, use \textbf{FIR} (finite impulse response) filters with appropriate order to achieve the desired frequency response. FIR filters (especially when applied in a forward-backward zero-phase manner) do not introduce phase distortions, which is crucial for preserving the shape of ERPs. They do, however, require more data points (a longer convolution kernel), which may introduce edge artifacts (one should ideally pad the data or use reflection at the edges). \textbf{IIR} (infinite impulse response) filters, like Butterworth, are more computationally efficient but can introduce non-linear phase unless applied zero-phase (which requires forward-backward filtering). If using IIR, always apply \texttt{filtfilt} (zero-phase). Modern toolboxes often default to optimized FIR designs; for example, EEGLAB and MNE default to Hamming-windowed FIR filters with specified transition bands~\cite{tadel2011brainstorm, tadel2019_, medani2023brainstorm}. Researchers should pay attention to the transition band---the range over which the filter goes from passband to stopband. Too narrow a transition can cause long filters (potentially cutting into data length) and ripples. For instance, MNE uses a default transition bandwidth of 0.5~Hz for low-pass, making very steep filters but very long (10s of data)~\cite{mnefiltering}. One can relax this (e.g., allow 2~Hz transition) to shorten the filter kernel.
\end{itemize}

\begin{bestpractice}
Use established defaults from known libraries unless you have specific needs. Report the cutoff frequencies and filter type in publications (e.g., ``Data were high-pass filtered at 0.1~Hz (zero-phase FIR, Hamming window, 1~Hz transition bandwidth) and low-pass filtered at 40~Hz (zero-phase FIR)''). This transparency aids reproducibility and interpretation~\cite{vandriel2019}.
\end{bestpractice}

\begin{itemize}[leftmargin=*]
  \item \textbf{Avoiding Filter Artifacts:} Be mindful of edge artifacts; filter convolution can cause edge transients. Mitigate by padding data or ignoring a margin after filtering. Also, check for filter ringing: a sharp transient in data (like an EOG spike) can cause oscillatory ripples due to the filter's impulse response. This is usually minor with low cutoffs, but if noticed, consider using a filter with a Kaiser window or a different design that minimizes ringing. Some pipelines prefer minimum-phase filters for real-time applications (which introduce phase shifts but less latency). For offline analysis focusing on timing, zero-phase is preferable~\cite{mnefiltering}.
\end{itemize}

Filtering is a challenging and balancing act: remove as much noise as possible without compromising brain signals. Lean toward conservative filters (wider passbands) if unsure, and always double-check that important features (ERP onsets, oscillation peaks) remain intact post-filtering~\cite{mnefiltering}. Modern recommendations strongly counsel against aggressive high-pass filtering, which can cause edge effects at lower cutoffs and longer recordings, or to use baseline correction to handle slow shifts instead of a high-pass filter. The combination of a gentle high-pass ($\sim$0.1--0.5~Hz) and a moderate low-pass (30--80~Hz, depending on needs) is a sensible default for many EEG studies and pipelines.

\subsection{Detecting and Repairing Bad Channels}
\label{sec:bad-channels}

During EEG data collection, some electrodes may be ``bad'', yielding no signal or excessively noisy signals due to high impedance, detachment, wires, or amplifier issues. Identifying and managing bad channels is crucial because a single bad channel can introduce artifacts (spikes or noise) that propagate when using an average reference and can also bias ICA or other multivariate analyses during preprocessing. The detection of the bad channels is often done in multiple steps.

\textbf{Expert Visual Inspection:} Traditionally, an expert would visually inspect time-series, power-spectrum, or topographic plots to spot channels with flatlines, outliers, or abnormally large noise. While still valuable, this approach does not scale well to high-density EEG and may require significant time and expertise for long recordings, and, in most cases, it is subjective.

\textbf{Using Automated Metrics:} Modern pipelines use quantitative criteria to flag bad channels~\cite{pernet2020_,islam2016}. Common metrics include, but are not limited to:

\begin{itemize}[leftmargin=*]
  \item \textbf{Spectrum deviations:} Some methods check if a channel's power spectrum deviates from the typical spectra of other channels (excessive 50~Hz noise or unusual 1/f shape could indicate a bad electrode).
  \item \textbf{Low variance or flatline:} If a channel's signal variance over some minutes is near zero (within system noise), it likely flatlined (electrode fell off or shorted). Many use a threshold such as ``flat for $>$5~seconds'' as a trigger~\cite{pernet2020_}.
  \item \textbf{High variance:} Channels with variance several standard deviations above the channel median might be wildly noisy.
  \item \textbf{Correlation with neighbors:} A channel that is poorly correlated with its neighbors could be problematic (since EEG channels near each other usually pick up similar signals). For example, EEGLAB provides the CleanRawData plugin ~\cite{cleanrawdata} (which defines a bad channel if its correlation with a robust re-reference of other channels is below a threshold, e.g., $r < 0.8$).
\end{itemize}

Often, multiple criteria are combined in popular software such as Autoreject~\cite{engemann2015}, FASTER~\cite{nolan2010_}, and ASR~\cite{blum2019}.

\textbf{Fixing bad channels: ``Repair or interpolation.''} Once bad channels are identified, the typical approach is to exclude them from analysis and, if topography needs to be preserved, interpolate them from neighboring channels. Interpolation (often using spherical splines or nearest-neighbor averaging) replaces the bad channel's data with a weighted average of surrounding channels, effectively ``filling in'' the missing data. This is useful for topographic visualization or for maintaining equal channel count across subjects, particularly for group-level analysis. If too many channels are bad (e.g., $>$10\% of channels), interpolation might degrade data.

\begin{bestpractice}
Use an automated bad-channel detection early in preprocessing. For example, the pipeline by Pernet et al.\ (2021) first removes bad channels using the \emph{clean\_rawdata} plugin with criteria: flatline $>$5~s or correlation $<$0.8~\cite{pernet2020_}. In that study, they then interpolated those channels from the remaining signals. Automated detection can flag many channels if the parameters are too strict, so review the list of marked bad channels; if it flags an entire lobe's electrodes, something might be off. Combining human oversight with automation is ideal: trust the algorithm to find obvious issues, but do a quick visual confirmation.
\end{bestpractice}

After channel repair, it is wise to recompute the average reference or whichever reference is used (since bad channels can bias referencing). Many pipelines actually exclude bad channels, perform referencing/ICA, then interpolate at the end.

\subsection{Reference Schemes: Average, REST, Laplacian, and Source Derivations}
\label{sec:references}

The reference of an EEG recording is the voltage against which all scalp electrodes are measured. Choosing an appropriate reference scheme is essential because EEG measures differences in potential; the reference affects the data's appearance but not the underlying brain activity. A poor reference can introduce artificial patterns (e.g., if a reference is noisy or active). Here we outline common referencing strategies and best practices:

\begin{itemize}[leftmargin=*]
  \item \textbf{Common Reference during Acquisition:} EEG amplifiers often record against a single reference electrode (e.g., one mastoid, Cz, FCz, or an average of mastoids). This is the \emph{recording reference}. However, after data collection, one can digitally re-reference the data to a new scheme to better represent the potential distribution.

  \item \textbf{Average Reference:} Perhaps the most widely used post-hoc reference. In an \emph{average reference}, the mean of all channels' voltages at each time point is subtracted from each channel. This sets the average potential across the head to zero at every moment. The average reference is a reasonable choice for high-density montages where electrodes cover the whole head, assuming that the net sum of source activity is captured and that averaging yields a zero reference (the ``reference-free'' concept). Average referencing can reduce the influence of a single noisy reference and often improves SNR when many channels are present. However, one must be cautious when using a low-density EEG average reference; for example, 20 electrodes may be biased if some regions are not covered~\cite{piontonachini2019_}.

  \item \textbf{REST (Reference Electrode Standardization Technique):} REST approximates a ``zero reference at infinity'' (a theoretical point at infinity where potential is zero) by computationally re-referencing the data using a head model and forward solution~\cite{nunez2010}. It uses a volume-conduction model (often a spherical head model) to estimate potentials at infinity from scalp data. In practice, REST can yield results similar to the average reference for many datasets, but is particularly useful when certain assumptions hold (e.g., neural sources sum to zero at infinity). Yao (2001)~\cite{yao2001} introduced REST as a means to reduce biases that an average reference might introduce when sources outside the scalp are present. REST requires knowing electrode coordinates and an estimate of head conductivity. It can be considered a more sophisticated average reference.

  \item \textbf{Laplacian or Current Source Density (CSD) Reference:} The Laplacian reference is not a reference to a single channel, but rather a transformation that computes the second spatial derivative of the voltage field at each electrode, effectively referencing each electrode to a weighted average of its neighbors (a local reference). This results in signals that emphasize local activity right under each electrode (since global common signals are subtracted out). The Laplacian (or CSD) can sharpen topographies and is very useful for spatially focal phenomena (e.g., distinguishing adjacent sources)~\cite{bitbrain2025}. It also attenuates volume-conducted far-field signals, acting as a spatial high-pass filter. The cost is that it may amplify noise if coverage is sparse. To compute the Laplacian, one needs electrode positions and typically assumes spline interpolation on the head surface. It is especially popular in EEG studies of oscillations or in Brain--Computer Interface applications, where one wants channel-specific activity.

  \item \textbf{Linked-Mastoids or Other Linked References:} In older practice, linking two electrodes (like two earlobes or mastoids) and using that as a reference was common (the assumption being that averaging two earlobes approximates a neutral point). This can be fine if those sites are relatively inactive. However, if the reference picks up brain signals (e.g., mastoids might pick up temporal lobe activity or neck muscles), they will be present (with the opposite sign) in all channels. Many modern studies avoid this in favor of average or REST.

  \item \textbf{Other Source Derivations:} Sometimes, a reference is chosen based on a specific electrode location. For example, Cz or the vertex is commonly used in some clinical montages. In bipolar recordings (such as in some sleep montages), each channel represents the voltage difference between two neighboring electrodes rather than relative to a single reference. All of these approaches attempt to place the reference at a site with minimal brain signal. However, there is no truly neutral reference location. That is why the average reference is widely adopted in research settings: it uses the mean of all electrodes as the reference, distributing reference effects across the scalp and reducing the influence of any single site, assuming adequate electrode coverage.
\end{itemize}

\begin{bestpractice}
For high-density EEG recordings with adequate scalp coverage, the common average reference is generally the most appropriate and robust choice for analysis and reporting. It distributes reference effects across electrodes and minimizes bias from any single site. When the goal is to enhance spatial specificity or reduce broadly distributed activity, a Laplacian (CSD) transform may be applied, as it provides a reference-independent spatial representation. Model-based approaches such as REST can also be useful in specific contexts.

Single-electrode references should be avoided when possible, as activity at the reference site propagates to all channels. Because re-referencing can be performed offline, it is best practice to preserve the original acquisition reference and apply transformations during preprocessing as needed.
\end{bestpractice}

\subsection{Artifact Correction Techniques: ICA Variants, GEDAI, and Beyond}
\label{sec:artifact-correction}

After filtering, bad-channel removal, and referencing, a major step is artifact correction---removing or suppressing artifacts in the remaining channels. Unlike simply discarding bad channels or time segments, artifact-correction methods aim to salvage data by isolating artifact contributions and subtracting or rejecting them. Here we cover the main leading approaches, including Independent Component Analysis (ICA) and its variants (the workhorse of EEG artifact removal), Signal-Space Projection (SSP/SSS) used in MEG, Artifact Subspace Reconstruction (ASR) for automated transient artifact removal, and newer developments like automated IC classification (e.g., ICLabel) and source-informed ICA methods like SOBI.

\subsubsection{Independent Component Analysis (ICA)}
\label{sec:ica}

Independent Component Analysis (ICA) is a blind source separation technique and a cornerstone of EEG artifact removal~\cite{blum2019,delorme2007,brainproductsICA}. It assumes that EEG signals are linear mixtures of independent neural and non-neural sources. ICA attempts to unmix these signals into statistically independent components (ICs) whose linear combination reconstructs the observed data. Ideally, some components represent artifacts (e.g., blinks, muscle activity, cardiac signals), while others reflect neural activity~\cite{uriguen2015}. Several ICA algorithms are used in EEG research:

\vspace{-1em} 

\begin{itemize}[leftmargin=*]
  \item \textbf{Infomax ICA:} Introduced by Bell \& Sejnowski and popularized in EEGLAB (\texttt{runica}). It maximizes output entropy (minimizes mutual information) and is widely used.
  \item \textbf{Extended Infomax:} An extension that can separate both super-Gaussian and sub-Gaussian sources, improving robustness.
  \item \textbf{Picard:} Accelerates the maximization of the Infomax likelihood by using a relative L-BFGS algorithm preconditioned with sparse Hessian approximations~\cite{ablin2018faster}.
  \item \textbf{FastICA:} Uses fixed-point iteration to maximize non-Gaussianity (e.g., kurtosis). It is computationally efficient but may be less stable in some EEG datasets.
  \item \textbf{JADE and SOBI:} Algorithms based on second-order statistics. SOBI (Second-Order Blind Identification) diagonalizes covariance matrices at multiple time lags and can be particularly effective for temporally structured or periodic artifacts such as ECG or eye blinks.
  \item \textbf{AMICA (Adaptive Mixture ICA):} A more advanced method that models mixtures of source distributions and often yields high-quality decompositions. Its main drawback is the substantial increase in computational time.
  \item \textbf{Adaptive and Source-Informed ICA:} Online or adaptive ICA methods perform decomposition in real time or over sliding windows, which is useful in brain--computer interface (BCI) applications. However, these approaches are computationally demanding and require careful implementation. When auxiliary recordings such as EOG or ECG are available, ICA can be guided using prior information. Constrained ICA (cICA) incorporates knowledge about expected correlations with artifact channels. Hybrid approaches may also combine regression and ICA, for example, regressing out EOG activity before applying ICA to the residual signals.
\end{itemize}

Alternative approaches have been explored, including:
\vspace{-1em} 
\begin{itemize}[leftmargin=*]
  \item \textbf{Canonical Correlation Analysis (CCA):} A BSS method that isolates sources by maximizing their temporal autocorrelation, separating highly autocorrelated neural signals from broadband, low-autocorrelation muscle (e.g., EOG, EMG)~\cite{declercq2006,somers2016}.
  \item \textbf{Principal Component Analysis (PCA):} Mainly for dimensionality reduction rather than true source separation.
  \item \textbf{Non-negative Matrix Factorization (NMF)} and related matrix factorization techniques have been tested for artifact removal.
\end{itemize}

Despite these developments, conventional ICA remains the most widely adopted method. It is well-validated, widely available in major toolboxes (EEGLAB, Brainstorm, MNE, FieldTrip), and has consistently proven effective at separating artifacts while preserving neural signals. The combination of ICA decomposition and targeted component rejection remains one of the most established approaches in EEG preprocessing.

\subsubsection{Application of the ICA on the Data}
\label{sec:ica-application}

Once ICA is applied to preprocessed EEG data (typically after high-pass filtering and re-referencing), components are inspected to determine whether they represent neural activity or artifacts. Artifact ICs are identified using multiple features:

\begin{itemize}[leftmargin=*, topsep=2pt, itemsep=2pt, parsep=0pt, partopsep=0pt]
  \item \textbf{Topography (scalp map):} For example, a blink IC shows strong frontal weights, horizontal eye movements produce left--right frontal patterns, and muscle ICs often have focal or edge-of-scalp distributions.
  \item \textbf{Time series and power spectrum:} Blink components show large, slow deflections with power concentrated at low frequencies (delta/theta), whereas muscle components exhibit broadband, high-frequency activity (often $>$20~Hz). Cardiac components show periodic spikes corresponding to the heart rate.
\end{itemize}

\enlargethispage{1\baselineskip} 

Correlating IC activations with recorded EOG or ECG channels can further confirm artifact components, as high correlations indicate shared variance with those physiological signals. The user then ``rejects'' those ICs, meaning they set their activations to zero and reconstruct the EEG from the remaining ICs. This effectively subtracts out the artifact contributions~\cite{uriguen2015}. Done correctly, ICA can remove blinks, eye movements, ECG signals, and some muscle artifacts without losing much of the brain signal. It has the advantage of preserving all time points (unlike trial rejection) and often allows keeping more data.

\textbf{Caveats:} ICA assumes linear mixing and statistically independent sources; these assumptions are approximate in EEG. It works best with high density and lots of data (so the ICA model can separate things). If only a few channels or short data are available, ICA might not separate well. Also, ICA does not ``know'' what is brain vs.\ artifact; it could split one artifact into several ICs or mix brain and artifact in one IC. Skill is needed to interpret ICs.

\subsubsection{Automated IC Classification (ICLabel)}
\label{sec:iclabel}

In addition to manual inspection, automated classifiers such as ICLabel can help identify artifact components~\cite{piontonachini2019_}. ICLabel is a machine-learning--based tool (implemented in EEGLAB) that classifies independent components into categories such as Brain, Eye, Muscle, Heart, Line Noise, Channel Noise, and Other. The classification is based on features including scalp topography, power spectrum, and activity patterns. ICLabel provides probabilistic labels for each component, which can support and standardize artifact rejection. However, automated classification should be used as an aid rather than a replacement for expert inspection, particularly in atypical datasets.

\begin{bestpractice}
Before running ICA, remove bad channels and reject segments with large, non-stereotyped artifacts (e.g., movement spikes), as extreme transients can dominate the decomposition. Applying a modest high-pass filter (e.g., 0.5--1~Hz) is commonly recommended for ICA training, since very slow drifts can violate ICA's stationarity assumptions and degrade component separation.

The number of independent components equals the effective data rank. If average referencing is used, the rank is reduced by one; this should be accounted for during ICA computation (most modern toolboxes handle this automatically).

After decomposition, carefully inspect and document the components that are removed. Reporting the typical number and type of rejected components (e.g., blink and eye-movement components identified by frontal topographies and EOG-correlated time courses) improves transparency and reproducibility. Finally, preserve the ICA weights and decomposition matrices. Saving these enables reproducibility, facilitates re-analysis, and allows inclusion in shared datasets (e.g., as BIDS derivatives).
\end{bestpractice}

\subsubsection{Other Artifact Removal Methods: Regression, SSP/SSS, ASR and GEDAI}
\label{sec:other-artifact-methods}

While ICA is widely used, several alternative or complementary approaches are available.

\begin{itemize}[leftmargin=*]
  \item \textbf{Regression-Based Artifact Removal:} Regression removes artifacts by modeling their contribution to EEG channels using a recorded reference signal (e.g., EOG or ECG). For each EEG channel, the artifact channel is used as a predictor in a linear model, and the estimated contribution is subtracted from the data. This approach is straightforward and effective for stereotyped artifacts such as eye blinks or cardiac activity, particularly when high-quality auxiliary recordings are available. However, regression assumes a linear relationship and may under- or over-correct the data. Neural activity correlated with the artifact channel can also be inadvertently removed. Adaptive filtering extends this approach by continuously updating regression coefficients, making it suitable for non-stationary artifacts or online applications. Regression is also commonly used to remove ballistocardiogram artifacts in simultaneous EEG--fMRI recordings. A detailed treatment of template-based approaches for gradient and pulse artifact correction in the EEG-fMRI context is provided in Section~\ref{sec:eegfmri}.

  \item \textbf{Signal-Space Projection (SSP) and Signal-Space Separation (SSS):} SSP removes artifacts by identifying their spatial pattern and projecting the data onto the orthogonal subspace. For example, blink epochs can be averaged to estimate an artifact topography, which is then removed via spatial projection. Conceptually, SSP is related to PCA-based subspace removal and is simpler than ICA but less flexible. SSS, also known as Maxwell filtering, is primarily used in MEG. It separates signals originating within the sensor array from external noise sources using physical modeling of magnetic fields. Although SSS does not directly apply to EEG, similar spatial filtering principles are conceptually related.

  \item \textbf{Artifact Subspace Reconstruction (ASR):} Artifact Subspace Reconstruction (ASR) is an automated method for suppressing transient, high-variance artifacts. It operates by comparing short-window covariance matrices to a baseline ``clean'' covariance estimate. When variance exceeds a predefined threshold, the affected subspace is reconstructed by removing outlier principal components. ASR is particularly effective for non-stationary noise bursts and movement artifacts, making it popular in mobile EEG and large datasets. Its aggressiveness is controlled by a burst threshold parameter: lower values remove more activity (risking overcorrection), whereas higher values are more conservative. ASR can operate online and does not require manual component selection. Careful selection and transparent reporting of such thresholds are essential, as overly aggressive settings risk removing neural signal, whereas conservative settings may leave residual artifacts.

  \item \textbf{Generalized Eigenvalue De-Artifacting Instrument (GEDAI):} A recent addition to EEG artifact correction methods is the Generalized Eigenvalue De-Artifacting Instrument (GEDAI)~\cite{ros2025}, introduced by Ros and colleagues in 2025. Unlike fully blind approaches such as ICA, GEDAI incorporates information from a forward model of the EEG leadfield to help distinguish neural activity from artifact-related components. The method uses generalized eigenvalue decomposition to contrast the covariance structure of the recorded data with a model-based covariance that reflects physiologically plausible brain sources. Components that do not align well with the forward model are attenuated, and the cleaned signal is reconstructed from the remaining subspace. GEDAI operates in an automated and unsupervised manner. It does not require auxiliary channels such as EOG or ECG, nor does it depend on manual component labeling. A data-driven criterion is used to determine the separation between signal and noise subspaces, which makes the approach suitable for large datasets and automated pipelines. Initial evaluations suggest that GEDAI performs competitively with ICA- and ASR-based pipelines, particularly in recordings with low signal-to-noise ratios or complex mixtures of artifacts. Because GEDAI relies on a forward model, its effectiveness depends on accurate sensor geometry and head modeling. Careful validation is therefore important when applying it to new datasets.
\end{itemize}

\subsubsection{Emerging and Complementary Approaches}
\label{sec:emerging-artifact}

Beyond established and model-informed approaches, several emerging strategies are being explored for EEG artifact correction.

\begin{itemize}[leftmargin=*]
  \item \textbf{Deep learning--based artifact removal} methods use neural networks to learn mappings from contaminated EEG to cleaned signals. Architectures such as denoising autoencoders, convolutional neural networks, and more recently diffusion models have been proposed to address complex and non-stationary artifacts. These approaches can outperform traditional ICA-, PCA-, or ASR-based pipelines in specific scenarios, such as gradient artifacts in simultaneous EEG--fMRI or highly non-linear noise patterns. However, they typically require large and well-characterized training datasets, and their generalizability across recording setups remains an active area of research~\cite{raj2025comprehensive}.
  \item \textbf{Transform-based and hybrid methods} include wavelet thresholding, empirical mode decomposition (EMD), singular spectrum analysis (SSA), and combinations of decomposition techniques such as EMD coupled with blind source separation. These approaches exploit time--frequency structure or intrinsic oscillatory modes to isolate artifact components. While widely studied in methodological literature, they are less commonly integrated into standard preprocessing toolboxes compared to ICA or ASR~\cite{jiang2019removal}.
\end{itemize}

\paragraph{Recommendations and Best Practices.}
Artifact correction should be systematic, transparent, and adapted to the specific characteristics of the dataset rather than relying on a single technique. ICA remains one of the central methods in EEG preprocessing because it can separate overlapping neural and artifact sources when applied carefully. Its performance improves when it is preceded by appropriate steps such as bad channel removal, moderate high-pass filtering, and rank-aware decomposition. More advanced ICA variants such as AMICA or SOBI, along with automated classifiers like ICLabel, can further improve consistency and scalability.

Automated subspace-based approaches provide useful complementary options. ASR is particularly effective for suppressing transient high-variance artifacts and is widely used in mobile EEG and large datasets where manual inspection is impractical. GEDAI represents a newer development that incorporates forward-model information to distinguish physiologically plausible brain activity from artifact-dominated components.

No single method is optimal for all artifact types. In practice, combining approaches---such as filtering, segment rejection, ICA-based component removal, and when appropriate ASR, GEDAI, or regression---often produces the most robust results. Method selection should be guided by the dominant artifact type, the recording setup, the available auxiliary signals, and the goals of the study.

Clear reporting of algorithms, parameter choices such as ICA rank, ASR thresholds, or modeling assumptions in GEDAI, and documentation of removed components or subspaces is essential for reproducibility. As practical starting points, the \texttt{clean\_rawdata} EEGLAB 
plugin~\cite{cleanrawdata} uses the following defaults: an ASR burst 
threshold of 20, a channel correlation threshold of 0.8, and a 
line-noise z-score criterion of 4. For ICA, sufficient data length 
is important for stable decomposition; a commonly applied guideline 
suggests at least $20 \times k^2$ data points, where $k$ is the number of 
channels~\cite{makeig2011ica}. These values should be treated as 
initial defaults and adjusted based on data characteristics. The overall aim is to preserve as much neural signal as possible while minimizing residual artifact in a principled and well-documented manner.

\begin{table}[htbp]
\centering
\caption{Common EEG artifact types and their typical removal approaches. Preferred methods reflect widely used practices in EEG preprocessing; alternative strategies may be appropriate depending on data quality, recording setup, and research objectives.}
\label{tab:artifact-methods}
\footnotesize
\renewcommand{\arraystretch}{1.1}
\begin{tabularx}{\textwidth}{@{} 
  >{\raggedright\arraybackslash}p{2.1cm} 
  >{\raggedright\arraybackslash}p{2.3cm} 
  >{\raggedright\arraybackslash}p{2.5cm} 
  >{\raggedright\arraybackslash}X @{}}
\toprule
\textbf{Artifact Type} & \textbf{Primary Method(s)} & \textbf{Alternative / Complementary Approaches} & \textbf{Notes} \\
\midrule
Eye blinks & ICA & Regression (EOG); SSP & Trial rejection is possible but reduces data. ICA typically preserves more neural signal. \\
\addlinespace[0.2em]
Eye movements & ICA & Regression (EOG) & Horizontal movements often produce left--right mirrored components. \\
\addlinespace[0.2em]
Muscle (EMG) & ICA (focal EMG comp.) & ASR (bursts); epoch rejection; GEDAI & Low-pass filtering reduces EMG but also attenuates high-frequency neural activity. \\
\addlinespace[0.2em]
Heartbeat (ECG) & ICA; Regression (ECG) & --- & Cardiac components show a periodic structure. In simultaneous EEG-fMRI, pulse artifact correction requires specialized template-based methods (see Section~\ref{sec:eegfmri}). \\
\addlinespace[0.2em]
Line noise (50/60~Hz) & Notch filtering; spectral methods & --- & Typically spatially widespread. ICA rarely isolates it effectively. \\
\addlinespace[0.2em]
Slow drift / sweat & High-pass filtering & Detrending & Prevention through proper skin preparation and stable temperature is critical. \\
\addlinespace[0.2em]
Motion artifacts & ASR & Segment rejection; motion-informed regression; GEDAI & Difficult to model. Prevention and quality control remain essential. \\
\addlinespace[0.2em]
Mixed or low SNR & ICA; GEDAI & ASR & Subspace-based approaches may be advantageous when multiple artifact types overlap. \\
\bottomrule
\end{tabularx}
\end{table}

\subsection{Simultaneous EEG-fMRI: Cardiac and Pulse Artifacts, Field Dependence, and Practical Guidance}
\label{sec:eegfmri}

As noted in Section~\ref{sec:noise-sources}, cardiac-related artifacts are especially pronounced in simultaneous EEG-fMRI recordings. The MRI environment introduces artifact sources that go beyond those addressed by the general correction framework described in Section~\ref{sec:artifact-correction}. Specifically, gradient artifact and pulse artifact require dedicated correction steps before standard methods such as ICA, ASR, or regression can be meaningfully applied to the residual EEG signal.

Cardiac-related artifacts in EEG recorded inside the MRI environment are commonly referred to as pulse artifact (PA) or ballistocardiogram (BCG) artifact. These arise from interactions between the cardiac cycle and the static magnetic field ($B_0$), producing voltages via multiple mechanisms. Pulsatile blood flow, an electrically conductive fluid, moving within $B_0$ induces voltages consistent with Hall effect principles, while heartbeat-synchronized micro-movements of electrodes and lead wires within $B_0$ generate additional induction artifacts. Pulsatile arterial movement can further induce voltage contamination in the recorded signal~\cite{warbrick2022}.

In contrast to gradient artifact (GA), which is periodic and precisely time-locked to slice acquisition, PA/BCG is quasi-periodic and non-stationary across time and channels. Pulse artifact correction using ECG-informed averaging was first described by Allen et al.\ (1998)~\cite{allen1998}. Average Artifact Subtraction (AAS) was subsequently introduced for gradient artifact removal in simultaneous EEG-fMRI recordings~\cite{allen2000}. Although alternative approaches such as optimal basis set (OBS) modeling have been proposed~\cite{niazy2005}, template-based correction remains widely used in practice~\cite{warbrick2022}. These template-based approaches are conceptually related to the regression-based artifact removal methods discussed in Section~\ref{sec:other-artifact-methods}: both estimate the artifact contribution from a reference signal or model and subtract it from the EEG data. AAS models gradient and pulse artifacts as repeating templates and subtracts them from the EEG. After GA and PA correction and downsampling, the residual EEG can be processed using standard pipelines (e.g., ICA) for removal of remaining artifacts.

\begin{figure}[htbp]
  \centering
  \includegraphics[width=0.8\textwidth]{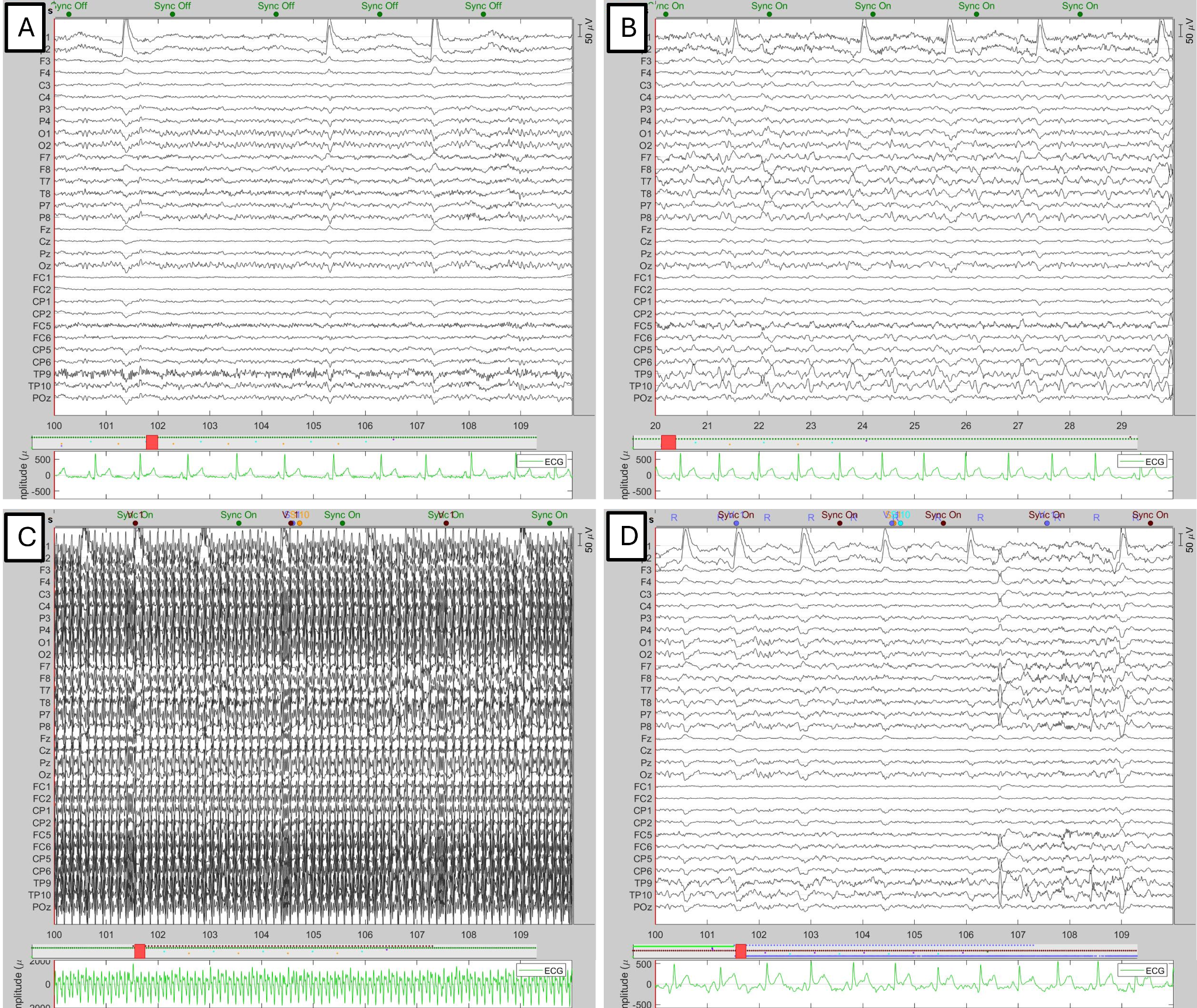}
  \caption{Representative EEG recordings across acquisition conditions. (A) Outside scanner room, (B) Inside scanner bore with no imaging sequence running (Scanner OFF), (C) Active imaging sequence (Scanner ON) showing high-amplitude gradient artifact, and (D) Final cleaned data after sequential gradient and pulse artifact correction. Reproduced from Razmara et al.\ (2026).}
  \label{fig:eegfmri}
\end{figure}

A representative example of the EEG signal across acquisition conditions is shown in Figure~\ref{fig:eegfmri}, demonstrating the magnitude of the gradient artifact during scanning and its attenuation following gradient and pulse artifact template-based correction~\cite{razmara2026eegfmri}.

PA/BCG amplitude depends on magnetic field strength. Artifact amplitude has been shown to increase with static field strength, making simultaneous EEG-fMRI more challenging at 3T and above compared to lower-field systems~\cite{warbrick2022}. This field dependence suggests that lower-field systems may offer a practical advantage for simultaneous EEG-fMRI by reducing artifact amplitude and simplifying preprocessing. For example, in recent feasibility studies at 0.55T, reduced BCG amplitude relative to high-field reports has been observed~\cite{razmara2026eegfmri,razmara2025fmri}, and this lower artifact amplitude facilitated effective denoising using synchronized AAS combined with ECG-informed template correction.

From a practical standpoint, EEG preprocessing inside the scanner should follow a structured sequence. High sampling rates (typically 5~kHz) are required to accurately capture GA transients~\cite{allen2000}. Precise synchronization between MRI clock and EEG acquisition minimizes residual artifact. GA correction should precede PA correction. Following GA removal and downsampling, PA correction should be performed using ECG-informed event detection with manual or semi-automatic verification to avoid errors introduced by misidentified R-peaks~\cite{warbrick2022,allen1998}.

Successful preprocessing should be validated by demonstrating preservation of physiological rhythms and task-evoked responses~\cite{warbrick2022,telesford2023}. For example, in 0.55T simultaneous EEG-fMRI recordings, alpha-band activity and occipital stimulus-locked steady-state visual responses were preserved after correction~\cite{razmara2026eegfmri,razmara2025fmri}.

The same principles extend beyond conventional BOLD EPI to real-time MRI (rtMRI) sequences used in speech research. Recent work has demonstrated speech production real-time MRI at 0.55T~\cite{lim2024,jihwan2026}, where rapid (millisecond-scale) gradient switching and continuous acquisition introduce structured induction artifacts analogous to those observed in fMRI-based EEG recordings.

\subsection{Organizing Data with BIDS-EEG and Maintaining Reproducibility}
\label{sec:bids-eeg}
Careful preprocessing does not by itself ensure scientific rigor. EEG pipelines involve many analytic decisions, including filter cutoff frequencies, reference schemes, bad-channel criteria, ICA rank selection, and artifact rejection thresholds. These choices influence downstream measures such as ERP amplitudes, spectral power, connectivity estimates, and decoding performance. Transparent documentation and standardized data organization are therefore essential components of best practice.

The Brain Imaging Data Structure (BIDS) provides a standardized framework for organizing neuroimaging datasets in a consistent and machine-readable format~\cite{pernet2019bids_}. The EEG extension, BIDS-EEG, adapts this framework specifically to electrophysiological recordings. Instead of lab-specific naming conventions or loosely documented metadata, BIDS enforces a structured directory hierarchy, standardized file names, and sidecar files that describe acquisition parameters and experimental events~\cite{gorgolewski2016}. This structure improves transparency, interoperability across software platforms, and long-term usability~\cite{pernetcobidas_,keil2014}.

A BIDS-EEG dataset is organized by subject and session, for example \texttt{sub-01/ses-01/eeg/}, and includes the raw EEG recording along with associated metadata. These typically include a JSON sidecar file specifying sampling rate, amplifier information, hardware filters, and recording reference; a \texttt{channels.tsv} file describing channel types and status; an \texttt{electrodes.tsv} file with electrode coordinates when available; a \texttt{coordsystem.json} file defining the coordinate system; and an \texttt{events.tsv} file documenting stimulus timing and condition labels. By separating signal data from metadata while maintaining explicit links, BIDS reduces ambiguity and supports automated processing.

An important feature of BIDS is the distinction between raw data and derivatives. Raw recordings should remain unchanged. Preprocessing outputs such as filtered data, re-referenced signals, ICA decompositions, artifact-corrected data, and time--frequency representations should be stored as derivatives. Each derivative should include metadata describing filter type and cutoff frequencies, reference scheme, ICA algorithm and rank, rejected components, artifact correction parameters, channel interpolation procedures, and software versions. This traceability makes it possible to reproduce analyses or compare alternative preprocessing strategies~\cite{wilkinson2016_,appelhoff2019}.

Reproducibility also depends on workflow management. Preprocessing steps should be scripted and version controlled. Software versions must be documented. When analyses involve stochastic procedures, such as ICA initialization or machine learning training, random seeds should be recorded. Automated pipelines reduce variability introduced by manual intervention and improve consistency across subjects and sessions. These practices align with broader recommendations for transparency in neuroimaging research~\cite{nichols2017_,poldrack2017_}.

The relevance of structured data management extends directly to the computational modeling approaches described in Sections~\ref{sec:computational-modeling} and~\ref{sec:machine-learning}. ERP estimation depends on clearly defined epoching and baseline procedures. Time--frequency analysis depends on documented filter settings and wavelet parameters. Connectivity measures are sensitive to referencing and spatial preprocessing. In machine learning applications, improper documentation of preprocessing steps can lead to data leakage or inflated performance estimates~\cite{varoquaux2017}. Organizing data and derivatives within the BIDS framework reduces hidden analytic flexibility and supports valid statistical inference.

BIDS-EEG also aligns with the FAIR data principles, which emphasize that scientific data should be findable, accessible, interoperable, and reusable~\cite{wilkinson2016_}. Public repositories increasingly require or encourage BIDS compliance. Even when datasets cannot be openly shared, internal adoption of BIDS improves collaboration, lab continuity, and methodological consistency.

From a practical perspective, EEG data should be converted to BIDS format soon after acquisition~\cite{gorgolewski2017bidsapps}. Preprocessing can then proceed within this standardized structure, and derivatives can be written back into the dataset hierarchy. Toolboxes such as MNE-Python~\cite{appelhoff2019}, EEGLAB, FieldTrip, and Brainstorm provide support for BIDS import and export, facilitating integration into automated workflows.

Structured data organization is therefore foundational to robust EEG research. Preservation of raw data, explicit documentation of preprocessing parameters, and standardized storage of derivatives ensure that statistical and machine learning analyses are reproducible, interpretable, and comparable across studies.

\section{Computational Modeling of EEG Data}
\label{sec:computational-modeling}

Even after careful preprocessing, the neural signals of interest remain difficult to isolate in EEG recordings. This limitation arises from the biophysical nature of EEG measurement. Because of volume conduction, each scalp electrode captures a spatially diffuse superposition of electrical activity from distributed cortical sources. By the time the signal reaches the sensor, task-related dynamics are inseparably mixed with ongoing background activity and physiological and environmental artifacts. Although preprocessing can remove some of these confounds, it cannot fully resolve the fundamental mixing problem inherent to scalp recording.

Therefore, extracting meaningful structures from EEG recordings requires computational modeling. Without principled approaches to disentangle overlapping sources and characterize task-relevant representations, the signal of interest remains obscured within high-dimensional data. Over the past decades, a diverse set of modeling strategies has been developed to address this challenge, ranging from classical univariate signal processing to multivariate decoding frameworks using modern artificial intelligence. In this chapter, we introduce some widely used foundational approaches for EEG analysis and examine how they enable inference of neural activity from mixed scalp recordings.

\subsection{Event-Related Potential Analysis}
\label{sec:erp}

Event-related potential (ERP) analysis remains one of the most established and interpretable approaches for studying task-evoked neural dynamics using EEG. ERPs are obtained by time-locking EEG signals to discrete events, segmenting the data into pre- and post-stimulus epochs, applying baseline correction using the pre-stimulus interval~\cite{pernetcobidas_,keil2014}, and averaging across repeated trials, under the assumption that stimulus-locked neural responses are temporally consistent, whereas unrelated activity varies randomly across trials. This process captures event-related neural responses relative to the ongoing background activity, enhances the task-relevant signal-to-noise ratio (SNR), and reveals characteristic voltage deflections that are associated with sensory, motor, and cognitive processes~\cite{luck2014}. ERP waveforms are estimated at selected electrodes or by averaging signals across a cluster of neighboring electrodes. ERP analysis provides a temporally resolved representation of neural activity, allowing researchers to examine component amplitudes, latencies, and scalp distributions as indices of underlying neural activity.

\begin{figure}[htbp]
  \centering
  \includegraphics[width=0.65\textwidth]{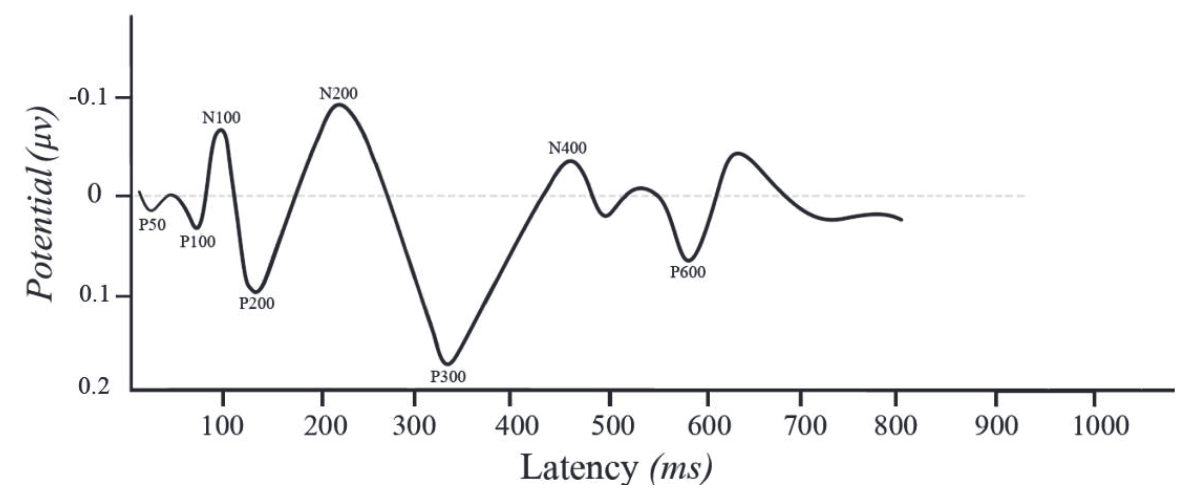}
  \caption{Example ERP waveform~\cite{srimaharaj2021}.}
  \label{fig:erp-waveform}
\end{figure}

ERP waveforms are characterized by a sequence of positive and negative voltage deflections that are described in terms of their polarity and latency. These deflections, commonly referred to as ERP components, are typically labeled using a letter-number convention. The letter (N/P) denotes polarity (negative/positive), and the accompanying number indicates the approximate peak latency in milliseconds (e.g., N100, P300) or the component's ordinal position within the waveform (e.g., P1, N2).

\begin{itemize}[leftmargin=*]
  \item \textbf{P50:} An early positive ERP deflection that typically peaks around 50~ms after stimulus onset, most prominently over central and fronto-central electrodes. It is commonly examined in sensory gating paradigms, where the P50 response to the second stimulus is attenuated relative to the first, reflecting inhibitory suppression of redundant input~\cite{light2003}. The degree of this attenuation is widely used as an index of sensory gating efficiency and has been shown to be altered in several neuropsychiatric conditions, particularly schizophrenia~\cite{potter2006}.

  \item \textbf{P100/P1:} A positive ERP component that emerges approximately 80--130~ms after stimulus onset, most prominently over occipital electrodes. It is commonly elicited in visual paradigms and is generally attributed to early sensory and perceptual processing within visual cortex. The amplitude of the P100 is sensitive to low-level stimulus characteristics such as luminance, contrast, and spatial frequency, and can also be modulated by spatial attention~\cite{hillyard1998,luck2014}.

  \item \textbf{N100/N1:} A negative ERP component that peaks around 80--120~ms after stimulus onset and is commonly observed over fronto-central electrodes in auditory paradigms. It is widely interpreted as reflecting early sensory processing within primary and secondary auditory cortex. N100 amplitude is sensitive to basic stimulus properties such as intensity, frequency, and onset characteristics, and is also modulated by attentional allocation~\cite{naatanen1987}.

  \item \textbf{P200/P2:} A positive ERP component that peaks around 150--250~ms after stimulus onset, commonly observed over fronto-central or central electrodes. It has been associated with early attentional processing, stimulus evaluation, and perceptual categorization. P200 amplitude is often modulated by stimulus salience, expectancy, and task relevance~\cite{crowley2004}.

  \item \textbf{N200/N2:} A negative ERP component that typically emerges from 200--350~ms after stimulus onset over fronto-central electrodes. It is commonly associated with cognitive control processes, including conflict monitoring, response inhibition, and mismatch detection. In tasks such as go/no-go or flanker paradigms, enhanced N200 amplitudes are often observed under conditions requiring increased inhibitory control~\cite{folstein2008}.

  \item \textbf{P300/P3:} A major positive ERP component related to attention allocation and context updating in working memory that peaks around 250--500~ms after stimulus onset over parietal electrodes. P300 amplitude is sensitive to stimulus probability, task relevance, and motivational salience. The component is often subdivided into the P3a, associated with novelty detection and fronto-central distribution, and the P3b, associated with task-relevant target processing and maximal over parietal sites~\cite{polich2007}.

  \item \textbf{N400:} A negative ERP component that typically peaks around 300--500~ms after stimulus onset and is most prominent over centro-parietal electrodes. It is classically elicited in language and semantic processing paradigms, where its amplitude is sensitive to the degree of semantic incongruity or expectancy violation~\cite{mcdaniel2026}. Beyond language, the N400 has also been observed in response to meaningful nonverbal stimuli, suggesting that it reflects a more general process of semantic integration or access to stored knowledge~\cite{kutas1980,kutas2011}.

  \item \textbf{P600:} A positive ERP component that emerges around 500--800~ms after stimulus onset over centro-parietal electrodes. It is most commonly observed in language paradigms involving syntactic anomalies or structural complexity. The component is widely interpreted as reflecting processes related to syntactic reanalysis, repair, or structural integration~\cite{kuperberg2007,osterhout1992}.
\end{itemize}

One major advantage of ERP analysis is its exceptional temporal resolution, which allows researchers to track neural processes on the order of milliseconds and examine the sequence of perceptual, cognitive, and affective processes as they unfold. Because ERP components are defined by their latency, polarity, and scalp distribution, they provide a relatively direct and interpretable framework for linking neural activity to specific stages of information processing.

A key limitation, however, is that ERP estimation relies on trial averaging and assumptions of temporal consistency across repetitions. Variability in response latency, overlapping neural processes, or non-phase-locked activity can attenuate or obscure component structure in the averaged waveform. Moreover, scalp-recorded ERPs reflect the summation of distributed cortical sources, limiting spatial specificity and complicating source interpretation.

\subsection{Time--Frequency Analysis}
\label{sec:time-frequency}

Time--frequency (TF) analysis provides a complementary approach to ERP analysis by characterizing the spectral dynamics of EEG signals over time. Rather than focusing exclusively on voltage deflections that are phase-locked to stimulus onset, time--frequency methods quantify changes in oscillatory power and phase across frequency bands. This distinction is important because neural oscillations are not necessarily phase-locked to stimulus onset, and averaging across trials may suppress or obscure non-phase-locked dynamics that nonetheless reflect meaningful task-related activity. TF analysis is typically performed by applying a spectral decomposition method (e.g., wavelet transform, short-time Fourier transform or Hilbert transform) to each trial to estimate time-resolved oscillatory power and phase across frequency bands. The resulting estimates are then baseline-corrected relative to a pre-stimulus interval to quantify stimulus-related changes in oscillatory activity. Finally, TF measures are averaged across trials to obtain event-related changes in oscillatory dynamics associated with the stimulus or task~\cite{cohen2014}.

Widely used methods for computing time--frequency representations of EEG signals are:

\begin{itemize}[leftmargin=*]
  \item \textbf{Fast Fourier Transform (FFT):} The FFT is an efficient algorithm for computing the discrete Fourier transform (DFT), which decomposes a time-domain signal into its constituent frequency components. In EEG analysis, FFT is commonly applied to estimate the power spectrum of neural signals, revealing how signal power is distributed across frequencies~\cite{oppenheim}.

  \item \textbf{Welch's Method (\texttt{pwelch}):} Welch's method is a spectral estimation technique that improves the stability of power spectral density (PSD) estimates by averaging spectra computed from multiple overlapping segments of the signal. In EEG analysis, Welch's method is widely used to obtain smoother and more reliable estimates of spectral power compared to a single FFT applied to the entire signal~\cite{welch1967}.

  \item \textbf{Short-Time Fourier Transform (STFT):} A Fourier-related transform method that characterizes how the spectral content of a signal evolves over time. Unlike the classical Fourier transform, which provides frequency information over the entire signal duration, the STFT applies the Fourier transform within a sliding temporal window, thereby preserving local temporal structure. A limitation of the STFT is the fixed trade-off between temporal and frequency resolution, determined by the chosen window length~\cite{cohen2014,owens1988}.

  \item \textbf{Wavelet Transform:} A time--frequency analysis technique that decomposes a signal into scaled and time-shifted versions of a localized waveform known as a wavelet. In practice, complex Morlet wavelets are commonly used in electrophysiological research to estimate time-resolved power and phase across frequency bands~\cite{cohen2014,tallonbaudry1999}.

  \item \textbf{Hilbert Transform:} A technique used to derive the analytic representation of a real-valued time series, enabling estimation of its instantaneous amplitude and phase. In EEG research, the signal is typically first band-pass filtered within a frequency range of interest, after which the Hilbert transform is applied to compute the complex analytic signal~\cite{cohen2014,boashash1992}.

  \item \textbf{Multitaper Analysis:} A spectral estimation technique designed to improve the reliability and stability of power estimates, particularly in noisy signals such as EEG. Rather than relying on a single window function, multitaper analysis applies a set of orthogonal tapers, typically discrete prolate spheroidal sequences (DPSS), to the same data segment and averages the resulting spectral estimates~\cite{mitra1999,thomson1982}.
\end{itemize}

A major strength of time--frequency analysis is its ability to characterize both phase-locked and non-phase-locked neural activity, thereby offering a richer description of task-related dynamics across oscillatory components. Studies have shown that oscillatory activities in specific frequency bands are associated with distinct functional roles.

\begin{itemize}[leftmargin=*]
  \item \textbf{Delta (0.5--4~Hz):} Delta-band activity is most prominent during deep sleep but is also observed in awake states under conditions involving sustained attention, motivational salience, or homeostatic regulation~\cite{harmony2013,buzsaki2004}.

  \item \textbf{Theta (4--8~Hz):} Theta oscillations are frequently linked to memory encoding and retrieval, cognitive control, and conflict monitoring. Frontal midline theta activity, in particular, is consistently observed during tasks requiring executive control~\cite{klimesch1999,cavanagh2014}.

  \item \textbf{Alpha (8--12~Hz):} Alpha activity is traditionally associated with relaxed wakefulness and posterior dominance during eyes-closed rest. Functionally, task-related alpha suppression (event-related desynchronization) is often interpreted as reflecting cortical activation, whereas alpha enhancement may index functional inhibition~\cite{pfurtscheller1999,klimesch2012}.

  \item \textbf{Beta (13--30~Hz):} Beta-band activity has been strongly associated with motor control and maintenance of the current sensorimotor or cognitive state. Beta desynchronization typically occurs during movement preparation and execution, followed by post-movement rebound~\cite{engel2010}.

  \item \textbf{Gamma (30--100+~Hz):} Gamma-band activity is often interpreted as reflecting local cortical processing and neuronal synchronization at fine temporal scales. However, high-frequency activity must be interpreted cautiously in EEG due to susceptibility to muscle artifacts~\cite{tallonbaudry1999}.
\end{itemize}

\subsection{Connectivity Analysis}
\label{sec:connectivity}

Beyond characterizing local neural activity, connectivity analysis quantifies statistical dependencies and directed interactions between spatially distributed brain regions. Connectivity measures are broadly categorized into functional and effective connectivity. Functional connectivity refers to statistical associations between signals, without implying directionality. Common functional connectivity metrics include correlation, coherence, phase-locking value (PLV), and phase-lag index (PLI)~\cite{bastos2015,friston2011}.

\begin{itemize}[leftmargin=*]
  \item \textbf{Correlation:} Correlation quantifies the linear relationship between two time-domain signals, typically using the Pearson correlation coefficient. Although simple and widely used, correlation is sensitive to common reference effects and volume conduction~\cite{bastos2015}.

  \item \textbf{Coherence:} Coherence measures frequency-specific linear coupling between two signals and is derived from the normalized cross-spectral density. Because coherence includes zero-lag interactions, it is susceptible to spurious coupling arising from field spread and common sources~\cite{bastos2015,nolte2004}.

  \item \textbf{Phase-Locking Value (PLV):} The phase-locking value quantifies the consistency of phase differences between two signals across trials or time. It isolates phase synchronization independent of amplitude fluctuations. However, PLV remains sensitive to zero-lag synchronization and volume conduction effects~\cite{lachaux1999}.

  \item \textbf{Phase-Lag Index (PLI):} The phase-lag index was developed to reduce the influence of volume conduction by measuring the consistency of non-zero phase lags between signals. By ignoring phase differences centered around zero, PLI reduces spurious connectivity caused by common sources~\cite{stam2007}.
\end{itemize}

Effective connectivity seeks to infer directed or causal interactions between neural signals. Methods such as Granger causality and directed transfer function (DTF) are commonly used.

\begin{itemize}[leftmargin=*]
  \item \textbf{Granger causality:} A directed connectivity measure based on multivariate autoregressive modeling. A signal $x$ is said to Granger-cause $y$ if past values of $x$ significantly improve prediction of $y$ beyond its own past~\cite{granger1969}. In EEG research, it is used to infer putative directional interactions, though interpretation depends on model assumptions and is sensitive to volume conduction~\cite{bastos2015}.

  \item \textbf{Directed Transfer Function (DTF):} A frequency-domain measure of effective connectivity derived from multivariate autoregressive models. It quantifies the directional influence between signals as a function of frequency~\cite{kaminski1991}. Reliable estimation requires appropriate model order selection and sufficient data length~\cite{blinowska2011}.
\end{itemize}

Connectivity analysis can be performed in sensor space or source space, and often within specific frequency bands. However, interpretation must be approached cautiously in EEG due to volume conduction and field spread, which can artificially inflate apparent coupling between nearby sensors.

\textbf{Source localization} aims to estimate the cortical generators of EEG signals by solving the electromagnetic inverse problem. This process typically involves two components: a forward model, which describes how neural sources project to the scalp based on head geometry and tissue conductivity, and an inverse solution, which estimates source activity from measured EEG data~\cite{baillet2001,medani2025editorial}.

\subsection{Multivariate Analysis, Decoding, and Modeling}
\label{sec:multivariate}

Univariate EEG analyses examine neural activity at individual electrodes, time points, or frequency bands independently. While such approaches provide interpretable summaries of localized effects, they do not account for the inherently distributed nature of neural representations. Multivariate analysis addresses this limitation by treating EEG recordings as high-dimensional patterns and analyzing the joint structure of activity across channels, time, and frequency. Rather than asking whether one channel differs between conditions, multivariate modeling asks whether the overall pattern of neural activity contains discriminative or predictive information.

The central premise of multivariate analysis is that distinct behaviors or cognitive functions are encoded as distinguishable patterns of distributed neural activity. These patterns can be conceptualized as trajectories evolving within a high-dimensional representational space defined by the joint activity of multiple neural features. The primary objective is to determine whether these trajectories form separable structures and to characterize the organization of the latent representations that give rise to observable EEG signals.

\begin{figure}[htbp]
  \centering
  \includegraphics[width=0.6\textwidth]{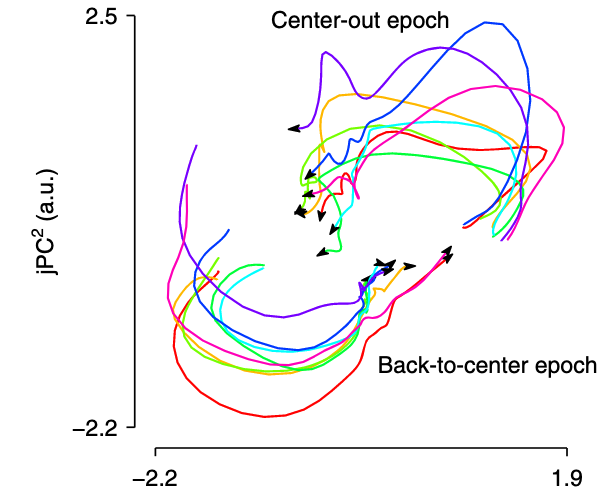}
  \caption{Example of low-dimensional latent neural trajectories during a reaching task (adapted from~\cite{kao2015}). Neural population activity projected onto jPC1 and jPC2 exhibits structured rotational dynamics.}
  \label{fig:neural-trajectories}
\end{figure}

Decoding approaches provide a practical framework for quantifying the discriminability between neural trajectories associated with different conditions. Distinctiveness may be evaluated using distance-based metrics, such as Euclidean or Mahalanobis distance, or by training classifiers such as support vector machines (SVM) or linear discriminant analysis (LDA) to separate activity patterns in feature space~\cite{blankertz2011,grootswagers2017}. In EEG research, time-resolved decoding has been widely used to reveal the temporal structure of sensory, motor, and cognitive processes~\cite{jeong2025decoding,jeong2023multivariate,kommineni2025}.

\begin{figure}[htbp]
  \centering
  \includegraphics[width=0.75\textwidth]{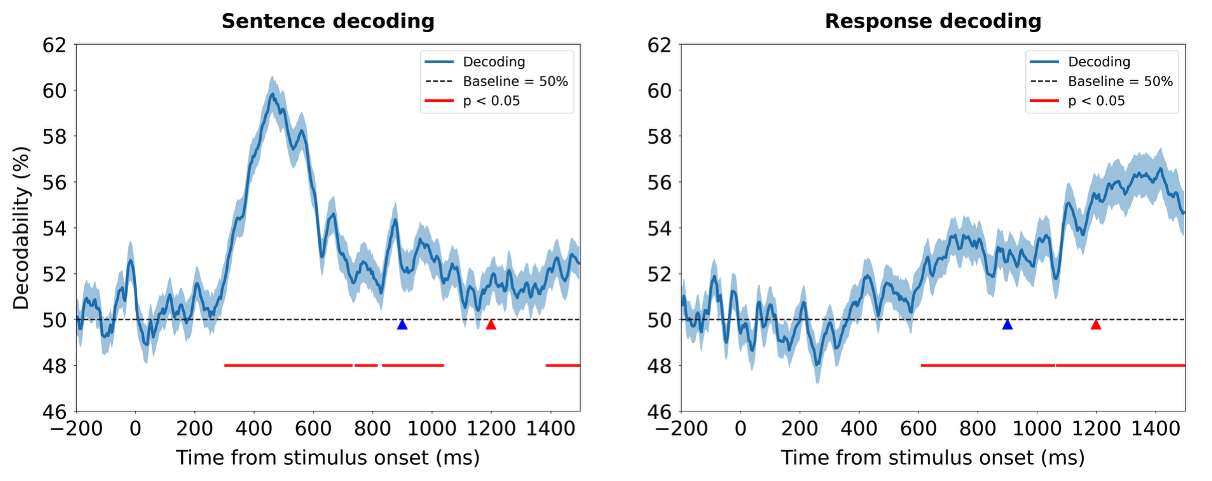}
  \caption{Example of time-resolved decoding during a cognitive task. Decodability over time indicates when neural representations differ between processing distinct sentences or producing different behavioral responses~\cite{jeong2025decoding}.}
  \label{fig:decoding}
\end{figure}

Representational similarity analysis (RSA) provides a complementary framework for examining the structure of neural representations by quantifying the similarity between multivariate activity patterns across experimental conditions. Neural responses are first estimated for each condition, and pairwise similarity or dissimilarity between activity patterns is computed using metrics such as correlation or Euclidean distance. The resulting representational dissimilarity matrix (RDM) summarizes the geometric organization of neural representations across conditions~\cite{kriegeskorte2008,cichy2014}.

Temporal generalization analysis extends time-resolved decoding by examining how neural representations generalize across time. In this approach, a decoder is trained at a specific time point and tested at other time points, producing a two-dimensional temporal generalization matrix that characterizes the stability and transformation of neural representations over time~\cite{king2014}.

Modeling approaches aim to explicitly characterize the latent structure underlying observed neural activity. State-space models, for example, assume that EEG signals arise from hidden dynamical states that evolve over time according to structured transition dynamics~\cite{durbin2012,cheung2010,galka2010}. Such models capture temporal continuity and internal organization of neural trajectories.

The inverted encoding model (IEM) is a modeling framework designed to reconstruct stimulus representations from patterns of neural activity. In this approach, neural responses are first modeled as weighted combinations of hypothetical feature-selective channels, each tuned to a particular stimulus dimension. After estimating this forward encoding model, the mapping is inverted to reconstruct the underlying stimulus representation~\cite{brouwer2009,brouwer2011}. In EEG studies, IEMs have been used to reconstruct sensory features and examine how stimulus representations evolve across time~\cite{brouwer2009,jeong2019,foster2016,foster2017}.

\subsection{Statistical Inferences in EEG Analysis}
\label{sec:statistics}

Computational modeling and signal processing methods allow researchers to extract patterns from EEG data; however, statistical inference is required to determine whether observed effects reflect reliable structure rather than sampling variability.

\textbf{Parametric and nonparametric tests:} Classical parametric tests assume specific distributional forms and are widely used when assumptions of normality and independence are approximately satisfied. Common examples include the $t$-test for comparing two conditions and analysis of variance (ANOVA) for factorial designs~\cite{student1908,fisher1992}.

When distributional assumptions are violated or when dealing with complex spatiotemporal EEG data, nonparametric approaches are often preferred. Permutation testing, for example, estimates the null distribution empirically by repeatedly shuffling condition labels and recomputing the statistic of interest~\cite{nichols2002}. Cluster-based permutation tests extend this framework to high-dimensional data by identifying clusters of adjacent significant samples in time, space, or frequency, and evaluating cluster-level statistics against a permutation-based null distribution~\cite{maris2007}.

\textbf{Multiple-comparison correction:} EEG analyses typically involve testing thousands of data points across electrodes and time, leading to inflated Type~I error rates if uncorrected. Widely used methods include Bonferroni correction~\cite{dunn1961}, false discovery rate (FDR) correction~\cite{benjamini1995}, cluster-based permutation testing~\cite{maris2007}, and permutation-based max-statistic correction~\cite{nichols2002}.

Robust EEG inference requires both appropriate statistical tests and principled control of multiple comparisons. Parametric tests provide interpretable hypothesis testing under standard assumptions, while permutation methods offer flexible and distribution-free alternatives. Error control strategies address the high-dimensional nature of EEG data. Together, these methods ensure that inferences drawn from sensor-level, time--frequency, or multivariate analyses reflect statistically reliable neural effects.

\section{Machine Learning for EEG: From Features to Classifiers and Beyond}
\label{sec:machine-learning}

While the previous section focused primarily on employing hand-crafted signal processing features (temporal, spectral and correlation based), in order to model neural processes underlying EEG signals, over the past decade deep learning models have brought about a paradigm shift from top-down feature modeling to bottom-up learning from raw data. Coupled with increases in scale of available data and compute power, deep learning models have provided impressive results in the domain of Natural Language Processing, Computer Vision and Speech Processing~\cite{devlin2019bert, baevski2020wav2vec, radford2021learning, radford2023robust, brown2020language}. The current section will delve into promising directions of deep learning models in the context of EEGcalongside some idiosyncratic characteristics of EEG that require further exploration of architectures in order to develop generalist EEG foundation models.

Owing to the inherent characteristics of the electroencephalogram (EEG)---such as high inter- and intra-subject variability, low signal-to-noise ratio (SNR), and heterogeneous recording conditions---deep learning models provide a natural fit for modeling complex neural dynamics. Recently, there has been an increasing interest in building generalized foundation models for the EEG modality. These models aim to provide state-of-the-art performance on a multitude of downstream tasks under zero-shot or low-resource settings.

Choosing an evaluation strategy that is appropriate for the training setting is equally important. For modeling paradigms where generalization is the primary goal, between-subjects cross-validation (leave-one-subject-out) must be prioritized. Conversely, in scenarios where personalization is crucial, within-subjects evaluation strategies should be utilized. To ensure reproducibility, leveraging standardized benchmarks like the Mother of All BCI Benchmarks (MOABB)~\cite{jayaram2018moabb} helps provide a fair comparative baseline.

\subsection{Classical Machine Learning Methods}
\label{sec:classical-ml}

Before the widespread adoption of deep learning, EEG classification pipelines relied heavily on rigorous feature engineering, spatial filtering, and dimensionality reduction as discussed in Section~\ref{sec:computational-modeling}.

\begin{itemize}[leftmargin=*]
  \item \textbf{Dimensionality Reduction and Artifact Removal:} Independent Component Analysis (ICA) is ubiquitously used to isolate and remove ocular and muscular artifacts from the raw EEG. Principal Component Analysis (PCA) is frequently employed to project high-dimensional electrode data into a lower-dimensional subspace, capturing the directions of maximum variance and reducing the computational burden for downstream classifiers.

  \item \textbf{Classical Classifiers:} Once features (e.g., band power, wavelets, or Common Spatial Patterns) are extracted, classical models are deployed. \textbf{Linear Discriminant Analysis (LDA)}~\cite{subasi2010eeg} and \textbf{Quadratic Discriminant Analysis (QDA)}~\cite{bhattacharyya2010performance} remain highly popular in Brain--Computer Interface (BCI) applications due to their low computational latency and robustness to limited data. \textbf{Support Vector Machines (SVM)}~\cite{guler2007multiclass, murugavel2016hierarchical}, often utilizing radial basis function (RBF) kernels, are highly effective in handling the non-linear boundaries of high-dimensional EEG feature spaces. Ensemble methods like \textbf{XGBoost} and \textbf{Random Forests} have also proven highly robust against the inherent noise and variance of EEG data.
\end{itemize}

\subsection{Deep Learning Models}
\label{sec:deep-learning}

Unlike classical machine learning models that have a feature extraction phase followed by a classification phase, deep learning models learn representations directly from raw or minimally preprocessed EEG signals. Often, CNN based models for EEG modelling use distinct filters for capturing the spatial and temporal associations followed by a late-fusion of these learnt features.

\begin{itemize}[leftmargin=*]
  \item \textbf{Convolution Based:} As EEG is a spatiotemporal signal, Convolutional Neural Networks (CNNs) are a well-suited architecture to capture the spatial correlations across electrodes.
    \begin{itemize}
      \item \textbf{EEGNet~\cite{Lawhern2016EEGNetAC}:} A compact CNN that utilizes depthwise and separable convolutions to encapsulate well-known EEG feature extraction concepts (like optimal spatial filtering) directly into its architecture. It is highly parameter-efficient and provides robust performance on motor-imagery, P300 and other BCI tasks.
      \item \textbf{SPaRCNet~\cite{jing2023development}:} Extensions and larger variants of CNN architectures designed for more complex clinical tasks. SPaRCNet is a deep CNN deployed in clinical settings to classify the ``ictal-interictal-injury continuum'' in ICU EEGs, demonstrating expert-level accuracy.
    \end{itemize}

  \item \textbf{Recurrent and Attention Based:} While Long Short-Term Memory (LSTM) networks were traditionally used to model the temporal progression of EEG signals, the field has largely shifted toward Transformer and attention-based architectures.
    \begin{itemize}
      \item Models like \textbf{ST-Transformer~\cite{song2021transformer}} (Spatiotemporal Transformer) and \textbf{EEGFormer} treat EEG channels and time-steps as sequences. They utilize self-attention mechanisms to capture long-range temporal dependencies and complex spatial synchronization across brain regions.
    \end{itemize}

  \item \textbf{Graph Based:} Because EEG electrodes are distributed across the 3D surface of the scalp, representing them as a non-Euclidean graph is highly intuitive. Graph Neural Networks (GNNs), such as Dynamical Graph CNNs (DGCNNs)~\cite{song2018eeg, xiao2025dynamical}, model the electrodes as nodes and dynamic functional connectivity as edges, allowing the network to learn complex topological brain networks.
\end{itemize}

\subsection{Self-Supervised Methods}
\label{sec:self-supervised}

Inspired by the success of Large Language Models (LLMs), self-supervised learning (SSL) foundation models leverage massive, unannotated EEG corpora (e.g., the TUH EEG Corpus~\cite{obeid2016temple}) to learn universal brain representations before fine-tuning on specific downstream tasks.

\begin{itemize}[leftmargin=*]
  \item \textbf{Contrastive Learning:} These methods learn representations by learning to embed similar EEG segments (e.g., augmented versions of the same window) closer in the latent space while pushing disparate segments farther apart.
    \begin{itemize}
      \item \textbf{BENDR~\cite{kostas2021bendr}:} Uses a transformer architecture combined with contrastive learning to model massive amounts of raw EEG data, generating universal features that adapt well across different subjects and hardware.
      \item \textbf{BIOT~\cite{yang2023biot}:} A Biosignal Transformer that tokenizes different biosignals into unified ``sentences,'' effectively handling mismatched channels and missing values across diverse datasets.
    \end{itemize}

  \item \textbf{Masked Reconstruction:} Similar to Masked Autoencoders (MAEs) in vision, these models mask a portion of the continuous EEG signal and train the network to reconstruct it.
    \begin{itemize}
      \item \textbf{CSBrain~\cite{zhou2025csbrain}:} A Cross-scale Spatiotemporal Brain foundation model that aggregates multi-scale features within localized temporal windows and brain regions. It uses structured sparse attention to capture dependencies and reconstruct masked segments.
      \item \textbf{CBraMod:~\cite{wang2024cbramod}} A ``criss-cross'' foundation model that utilizes a criss-cross transformer backbone. It models spatial and temporal dependencies separately through parallel attention mechanisms, employing patch-based masked EEG reconstruction.
      \item \textbf{S4-based Models~\cite{kommineni2024knowledge}:} Utilizing Structured State Space Sequence (S4)~\cite{gu2021efficiently} models to handle exceptionally long temporal dependencies in continuous EEG without the quadratic scaling bottleneck of traditional transformers. In addition to handling long sequences, S4 models are parameter efficient compared to transformers and are a more natural analogue for handling continuous time series signals such as EEG.
    \end{itemize}
\end{itemize}

\subsection{Comparative Analysis: The Discrepancy Between DL and SSL Performance}
\label{sec:dl-vs-ssl}

A prevailing question in modern EEG analysis is: \emph{Why are compact deep learning models (trained with a fraction of the parameter counts of massive SSL foundation models) able to provide highly competitive performance to foundation models on a multitude of tasks?} This observed discrepancy stems from several core characteristics of EEG data and model architecture:

\begin{enumerate}[leftmargin=*]
  \item \textbf{Strong Domain-Specific Inductive Biases:} Small DL models like EEGNet are highly optimized for EEG. They employ specific architectural choices---such as depthwise convolutions---that mathematically mirror established neuroscience techniques like Common Spatial Patterns (CSP). This strong inductive bias allows the model to learn efficiently from minimal data. Conversely, massive SSL transformers often use ``scale-agnostic'' dense modeling. Because they assume very little about the data's structure, they require exponentially more data to ``learn'' the spatial and temporal relationships that smaller models have hardcoded into their architecture.

  \item \textbf{The Nature of Signal-to-Noise Ratio (SNR):} Large SSL models borrow masked reconstruction and autoregressive objectives from natural language processing (NLP). However, language has a strict, universal semantic grammar. EEG data is highly stochastic, subjective, and plagued by low SNR~\cite{avramidis2025neural}. When an SSL model attempts to reconstruct masked EEG signals, a massive portion of its parameter capacity may be wasted memorizing intrinsic physiological noise and artifacts rather than learning meaningful cognitive representations.

  \item \textbf{Cross-Dataset Heterogeneity:} Foundation models attempt to build universal representations across datasets with different electrode montages, sampling rates, amplifier types, and reference schemes. The mathematical complexity of aligning these heterogeneous distributions limits the ``sharpness'' of the learned features. A compact DL model trained on a single dataset operates within a highly constrained, narrow distribution, allowing it to easily overfit to the exact predictive features needed for that specific task.
\end{enumerate}

\subsection{Evaluation Design, Data Leakage, and Performance Assessment}
\label{sec:ml-evaluation}

Choosing an appropriate evaluation strategy is as important as choosing the model itself. A major source of overoptimistic results in EEG machine learning is data leakage, where information from the test set inadvertently influences training~\cite{varoquaux2017}.

EEG data is particularly susceptible to leakage because trials recorded close in time share slow drifts, impedance fluctuations, and background brain state. If adjacent trials appear in both training and test sets, models can exploit this shared structure instead of learning generalizable neural features. This issue is especially severe in within-subject settings, where data are drawn from the same individual and temporal autocorrelation between trials is high. To mitigate this, data should be split along recording boundaries (sessions, blocks, or subjects). For cross-subject generalization, leave-one-subject-out cross-validation remains standard, while within-subject evaluation should use block-wise or session-wise splits rather than random trial-level shuffling~\cite{varoquaux2017}.

Feature or channel selection performed before cross-validation introduces leakage and must instead be performed within each training fold. Similarly, hyperparameter tuning on the test set produces biased estimates. Nested cross-validation or a held-out validation set should be used~\cite{cawley2010}.

Data augmentation can partially compensate for limited dataset size. Common approaches include temporal jittering, additive noise, and frequency-domain perturbations~\cite{rommel2022}. However, gains depend on the task and dataset and should be validated empirically. Augmentation used to expand the training set must be derived only from training samples. Test-time augmentation, where predictions are aggregated over transformed copies of a test sample, is a separate and valid practice.

Finally, reporting only classification accuracy is insufficient, particularly for imbalanced datasets (e.g., P300 paradigms). Metrics such as balanced accuracy, AUC, Cohen's kappa, or F1-score provide a more reliable assessment. Standardized benchmarks such as MOABB facilitate reproducible and comparable evaluation across studies~\cite{jayaram2018}.



\begin{thebibliography}{112}

\bibitem{bitbrain2025}
EEG Artifacts: Types, Detection, and Removal Techniques.
\emph{Bitbrain}, 2025.
\url{https://www.bitbrain.com/blog/eeg-artifacts}

\bibitem{brainproducts2022}
Brain Products.
Getting to know EEG artifacts and how to handle them in BrainVision Analyzer 2.
\emph{Brain Products Press Release}, 2022.

\bibitem{brainlatam2026}
How to handle EEG artifacts?
\emph{Brain Latam}, 2026.

\bibitem{blum2019}
S.~Blum, N.~S.~J.~Jacobsen, M.~G.~Bleichner, and S.~Debener.
A Riemannian modification of artifact Subspace Reconstruction for EEG artifact handling.
\emph{Front.\ Hum.\ Neurosci.}, 13:141, 2019.
\url{https://doi.org/10.3389/fnhum.2019.00141}

\bibitem{mnefiltering}
Background information on filtering --- MNE 0.23.4 documentation.
\url{https://mne.tools/0.23/auto_tutorials/preprocessing/25_background_filtering.html}

\bibitem{acunzo2012}
D.~J.~Acunzo, G.~Mackenzie, and M.~C.~W.~van~Rossum.
Systematic biases in early ERP and ERF components as a result of high-pass filtering.
\emph{J.\ Neurosci.\ Methods}, 209:212--218, 2012.

\bibitem{widmann2015}
A.~Widmann, E.~Schr\"oger, and B.~Maess.
Digital filter design for electrophysiological data---a practical approach.
\emph{J.\ Neurosci.\ Methods}, 250:34--46, 2015.

\bibitem{cleanline}
T.~Mullen.
\emph{CleanLine: EEGLAB plugin for removal of line noise}.
Neuroimaging Informatics Tools and Resources Clearinghouse (NITRC), 2012.
Available at: \url{https://eeglab.org/plugins/cleanline/}

\bibitem{mitra2008}
P.~Mitra and H.~Bokil.
\emph{Observed Brain Dynamics}.
Oxford University Press, 2008.

\bibitem{leske2019}
S.~Leske and S.~S.~Dalal.
Reducing power line noise in EEG and MEG data via spectrum interpolation.
\emph{NeuroImage}, 189:763--776, 2019.

\bibitem{dechevigne2019}
A.~de~Cheveign\'e.
ZapLine: a simple and effective method to remove power line artifacts.
\emph{bioRxiv}, 2019.

\bibitem{tadel2011brainstorm}
F.~Tadel, S.~Baillet, J.~C.~Mosher, D.~Pantazis, and R.~M.~Leahy.
Brainstorm: a user-friendly application for MEG/EEG analysis.
\emph{Comput.\ Intell.\ Neurosci.}, 2011:879716, 2011.

\bibitem{oostenveld2011_}
R.~Oostenveld, P.~Fries, E.~Maris, and J.-M.~Schoffelen.
FieldTrip: Open source software for advanced analysis of MEG, EEG, and invasive electrophysiological data.
\emph{Comput.\ Intell.\ Neurosci.}, 2011:156869, 2011.

\bibitem{gramfort2013_}
A.~Gramfort, M.~Luessi, E.~Larson, D.~A.~Engemann, D.~Strohmeier, C.~Brodbeck, et~al.
MEG and EEG data analysis with MNE-Python.
\emph{Front.\ Neurosci.}, 7:267, 2013.

\bibitem{tadel2019_}
F.~Tadel, E.~Bock, G.~Niso, J.~C.~Mosher, M.~Cousineau, D.~Pantazis, et~al.
MEG/EEG Group Analysis With Brainstorm.
\emph{Front.\ Neurosci.}, 13:76, 2019.

\bibitem{medani2023brainstorm}
T.~Medani, J.~Garcia-Prieto, F.~Tadel, M.~Antonakakis, T.~Erdbr\"ugger, M.~H\"oltershinken, et~al.
Brainstorm-DUNEuro: An integrated and user-friendly Finite Element Method for modeling electromagnetic brain activity.
\emph{NeuroImage}, 267:119851, 2023.

\bibitem{vandriel2019}
J.~van~Driel, C.~N.~L.~Olivers, and J.~J.~Fahrenfort.
High-pass filtering artifacts in multivariate classification of neural time series data.
\emph{bioRxiv}, 2019.

\bibitem{pernet2020_}
C.~R.~Pernet, R.~Martinez-Cancino, D.~Truong, S.~Makeig, and A.~Delorme.
From BIDS-formatted EEG data to sensor-space group results: A fully reproducible workflow with EEGLAB and LIMO EEG.
\emph{Front.\ Neurosci.}, 14:610388, 2020.

\bibitem{islam2016}
M.~K.~Islam, A.~Rastegarnia, and Z.~Yang.
Methods for artifact detection and removal from scalp EEG: A review.
\emph{Neurophysiol.\ Clin.}, 46:287--305, 2016.


\bibitem{cleanrawdata}
EEGLAB.
\emph{clean\_rawdata EEGLAB plugin}.
Available at: \url{https://eeglab.org/plugins/clean_rawdata/}

\bibitem{engemann2015}
D.~A.~Engemann and A.~Gramfort.
Automated model selection in covariance estimation and spatial whitening of MEG and EEG signals.
\emph{NeuroImage}, 108:328--342, 2015.

\bibitem{nolan2010_}
H.~Nolan, R.~Whelan, and R.~B.~Reilly.
FASTER: Fully Automated Statistical Thresholding for EEG artifact Rejection.
\emph{J.\ Neurosci.\ Methods}, 192:152--162, 2010.

\bibitem{piontonachini2019_}
L.~Pion-Tonachini, K.~Kreutz-Delgado, and S.~Makeig.
ICLabel: An automated electroencephalographic independent component classifier, dataset, and website.
\emph{NeuroImage}, 198:181--197, 2019.

\bibitem{nunez2010}
P.~L.~Nunez.
REST: a good idea but not the gold standard.
\emph{Clin.\ Neurophysiol.}, 121:2177--2180, 2010.

\bibitem{yao2001}
D.~Yao.
A method to standardize a reference of scalp EEG recordings to a point at infinity.
\emph{Physiol.\ Meas.}, 22:693--711, 2001.

\bibitem{delorme2007}
A.~Delorme, T.~Sejnowski, and S.~Makeig.
Enhanced detection of artifacts in EEG data using higher-order statistics and independent component analysis.
\emph{NeuroImage}, 34:1443--1449, 2007.

\bibitem{brainproductsICA}
Brain Products.
Independent Component Analysis (ICA) --- demystified.
\emph{Brain Products Press Release}, 2014.

\bibitem{uriguen2015}
J.~A.~Urig\"uen and B.~Garcia-Zapirain.
EEG artifact removal---state-of-the-art and guidelines.
\emph{J.\ Neural Eng.}, 12:031001, 2015.

\bibitem{ablin2018faster}
P.~Ablin, J.-F.~Cardoso, and A.~Gramfort.
Faster independent component analysis by preconditioning with {Hessian} approximations.
\emph{IEEE Trans.\ Signal Process.}, 66(15):4040--4049, 2018.

\bibitem{declercq2006}
W.~De~Clercq, A.~Vergult, B.~Vanrumste, W.~Van~Paesschen, and S.~Van~Huffel.
Canonical correlation analysis applied to remove muscle artifacts from the electroencephalogram.
\emph{IEEE Trans.\ Biomed.\ Eng.}, 53:2583--2587, 2006.

\bibitem{somers2016}
B.~Somers and A.~Bertrand.
Removal of eye blink artifacts in wireless EEG sensor networks using reduced-bandwidth canonical correlation analysis.
\emph{J.\ Neural Eng.}, 13:066008, 2016.

\bibitem{ros2025}
T.~Ros, V.~F\'erat, Y.~Huang, C.~Colangelo, S.~M.~Kia, T.~Wolfers, et~al.
Return of the GEDAI: Unsupervised EEG Denoising based on Leadfield Filtering.
\emph{bioRxiv}, 2025.

\bibitem{raj2025comprehensive}
V.~A.~Raj, T.~Parupudi, A.~Thalengala, and S.~G.~Nayak.
A comprehensive review of deep learning models for denoising EEG signals: challenges, advances, and future directions.
\emph{Discov.\ Appl.\ Sci.}, 7(11):1268, 2025.

\bibitem{jiang2019removal}
X.~Jiang, G.-B.~Bian, and Z.~Tian.
Removal of artifacts from {EEG} signals: a review.
\emph{Sensors}, 19(5):987, 2019.

\bibitem{makeig2011ica}
S.~Makeig and J.~Onton.
ERP features and EEG dynamics: An ICA perspective.
\emph{The Oxford Handbook of Event-Related Potential Components},
pp.~51--86. Oxford University Press, 2011.

\bibitem{warbrick2022}
T.~Warbrick.
Simultaneous EEG-fMRI: What have we learned and what does the future hold?
\emph{Sensors}, 22:2262, 2022.

\bibitem{allen1998}
P.~J.~Allen, G.~Polizzi, K.~Krakow, D.~R.~Fish, and L.~Lemieux.
Identification of EEG events in the MR scanner: the problem of pulse artifact and a method for its subtraction.
\emph{NeuroImage}, 8:229--239, 1998.

\bibitem{allen2000}
P.~J.~Allen, O.~Josephs, and R.~Turner.
A method for removing imaging artifacts from continuous EEG recorded during functional MRI.
\emph{NeuroImage}, 12:230--239, 2000.

\bibitem{niazy2005}
R.~K.~Niazy, C.~F.~Beckmann, G.~D.~Iannetti, J.~M.~Brady, and S.~M.~Smith.
Removal of FMRI environment artifacts from EEG data using optimal basis sets.
\emph{NeuroImage}, 28:720--737, 2005.

\bibitem{razmara2026eegfmri}
P.~Razmara, T.~Medani, M.~A.~Sisara, A.~A.~Joshi, R.~Chen, W.~Jeong, et~al.
Feasibility of simultaneous EEG-fMRI at 0.55T: Recording, denoising, and functional mapping.
\emph{arXiv:2602.13489}, 2026.

\bibitem{razmara2025fmri}
P.~Razmara, T.~Medani, A.~A.~Joshi, M.~A.~Sisara, Y.~Tian, S.~X.~Cui, et~al.
A feasibility study of task-based fMRI at 0.55T.
\emph{arXiv:2505.20568}, 2025.

\bibitem{telesford2023}
Q.~K.~Telesford, E.~Gonzalez-Moreira, T.~Xu, Y.~Tian, S.~J.~Colcombe, J.~Cloud, et~al.
An open-access dataset of naturalistic viewing using simultaneous EEG-fMRI.
\emph{Sci.\ Data}, 10:554, 2023.

\bibitem{lim2024}
Y.~Lim, P.~Kumar, and K.~S.~Nayak.
Speech production real-time MRI at 0.55T.
\emph{Magn.\ Reson.\ Med.}, 91:337--343, 2024.

\bibitem{jihwan2026}
J.~Lee, P.~Razmara, K.~Higa, S.~Fabus, A.~Kommineni, H.~Hothi, et~al.
An approach to simultaneous acquisition of real-time MRI video, EEG, and surface EMG for articulatory, brain, and muscle activity during speech production.
\emph{arXiv:2603.04840}, 2026.

\bibitem{pernet2019bids_}
C.~R.~Pernet, S.~Appelhoff, K.~J.~Gorgolewski, G.~Flandin, C.~Phillips, A.~Delorme, et~al.
EEG-BIDS, an extension to the brain imaging data structure for electroencephalography.
\emph{Sci.\ Data}, 6:103, 2019.

\bibitem{gorgolewski2016}
K.~J.~Gorgolewski, T.~Auer, V.~D.~Calhoun, R.~C.~Craddock, S.~Das, E.~P.~Duff, et~al.
The brain imaging data structure, a format for organizing and describing outputs of neuroimaging experiments.
\emph{Sci.\ Data}, 3:160044, 2016.

\bibitem{pernetcobidas_}
C.~Pernet, M.~I.~Garrido, A.~Gramfort, N.~Maurits, C.~M.~Michel, E.~Pang, et~al.
Issues and recommendations from the OHBM COBIDAS MEEG committee for reproducible EEG and MEG research.
\emph{Nat.\ Neurosci.}, 23:1473--1483, 2020.

\bibitem{keil2014}
A.~Keil, S.~Debener, G.~Gratton, M.~Jungh\"ofer, E.~S.~Kappenman, S.~J.~Luck, et~al.
Committee report: publication guidelines and recommendations for studies using electroencephalography and magnetoencephalography.
\emph{Psychophysiology}, 51:1--21, 2014.

\bibitem{wilkinson2016_}
M.~D.~Wilkinson, M.~Dumontier, I.~J.~J.~Aalbersberg, G.~Appleton, M.~Axton, A.~Baak, et~al.
The FAIR Guiding Principles for scientific data management and stewardship.
\emph{Sci.\ Data}, 3:160018, 2016.

\bibitem{appelhoff2019}
S.~Appelhoff, M.~Sanderson, T.~L.~Brooks, M.~van~Vliet, R.~Quentin, C.~Holdgraf, et~al.
MNE-BIDS: Organizing electrophysiological data into the BIDS format and facilitating their analysis.
\emph{J.\ Open Source Softw.}, 4:1896, 2019.

\bibitem{nichols2017_}
T.~E.~Nichols, S.~Das, S.~B.~Eickhoff, A.~C.~Evans, T.~Glatard, M.~Hanke, et~al.
Best practices in data analysis and sharing in neuroimaging using MRI.
\emph{Nat.\ Neurosci.}, 20:299--303, 2017.

\bibitem{poldrack2017_}
R.~A.~Poldrack, C.~I.~Baker, J.~Durnez, K.~J.~Gorgolewski, P.~M.~Matthews, M.~R.~Munaf\`o, et~al.
Scanning the horizon: towards transparent and reproducible neuroimaging research.
\emph{Nat.\ Rev.\ Neurosci.}, 18:115--126, 2017.

\bibitem{varoquaux2017}
G.~Varoquaux, P.~R.~Raamana, D.~A.~Engemann, A.~Hoyos-Idrobo, Y.~Schwartz, and B.~Thirion.
Assessing and tuning brain decoders: Cross-validation, caveats, and guidelines.
\emph{NeuroImage}, 145:166--179, 2017.

\bibitem{gorgolewski2017bidsapps}
K.~J.~Gorgolewski, F.~Alfaro-Almagro, T.~Auer, P.~Bellec, M.~Capot\u{a}, M.~M.~Chakravarty, et~al.
BIDS apps: Improving ease of use, accessibility, and reproducibility of neuroimaging data analysis methods.
\emph{PLoS Comput.\ Biol.}, 13:e1005209, 2017.

\bibitem{luck2014}
S.~J.~Luck.
\emph{An Introduction to the Event-Related Potential Technique}.
MIT Press, 2nd edition, 2014.

\bibitem{srimaharaj2021}
W.~Srimaharaj and R.~Chaisricharoen.
A novel processing model for P300 brainwaves detection.
\emph{J.\ Web Eng.}, 2021.

\bibitem{light2003}
G.~A.~Light and D.~L.~Braff.
Sensory gating deficits in schizophrenia: can we parse the effects of medication, nicotine use, and changes in clinical status?
\emph{Clin.\ Neurosci.\ Res.}, 3:47--54, 2003.

\bibitem{potter2006}
D.~Potter, A.~Summerfelt, J.~Gold, and R.~W.~Buchanan.
Review of clinical correlates of P50 sensory gating abnormalities in patients with schizophrenia.
\emph{Schizophr.\ Bull.}, 32:692--700, 2006.

\bibitem{hillyard1998}
S.~A.~Hillyard and L.~Anllo-Vento.
Event-related brain potentials in the study of visual selective attention.
\emph{Proc.\ Natl.\ Acad.\ Sci.\ USA}, 95:781--787, 1998.

\bibitem{naatanen1987}
R.~N\"a\"at\"anen and T.~Picton.
The N1 wave of the human electric and magnetic response to sound: a review and an analysis of the component structure.
\emph{Psychophysiology}, 24:375--425, 1987.

\bibitem{crowley2004}
K.~E.~Crowley and I.~M.~Colrain.
A review of the evidence for P2 being an independent component process: age, sleep and modality.
\emph{Clin.\ Neurophysiol.}, 115:732--744, 2004.

\bibitem{folstein2008}
J.~R.~Folstein and C.~Van~Petten.
Influence of cognitive control and mismatch on the N2 component of the ERP: a review.
\emph{Psychophysiology}, 45:152--170, 2008.

\bibitem{polich2007}
J.~Polich.
Updating P300: an integrative theory of P3a and P3b.
\emph{Clin.\ Neurophysiol.}, 118:2128--2148, 2007.

\bibitem{mcdaniel2026}
C.~McDaniel, M.~L.~Hughes, T.~Medani, W.~Jeong, T.~A.~McGee, A.~Kommineni, et~al.
Neural evidence of disrupted self-referential processing in suicidal depression.
\emph{J.\ Affect.\ Disord.}, 121817, 2026.

\bibitem{kutas1980}
M.~Kutas and S.~A.~Hillyard.
Reading senseless sentences: brain potentials reflect semantic incongruity.
\emph{Science}, 207:203--205, 1980.

\bibitem{kutas2011}
M.~Kutas and K.~D.~Federmeier.
Thirty years and counting: finding meaning in the N400 component of the event-related brain potential (ERP).
\emph{Annu.\ Rev.\ Psychol.}, 62:621--647, 2011.

\bibitem{kuperberg2007}
G.~R.~Kuperberg.
Neural mechanisms of language comprehension: challenges to syntax.
\emph{Brain Res.}, 1146:23--49, 2007.

\bibitem{osterhout1992}
L.~Osterhout and P.~J.~Holcomb.
Event-related brain potentials elicited by syntactic anomaly.
\emph{J.\ Mem.\ Lang.}, 31:785--806, 1992.

\bibitem{cohen2014}
M.~X.~Cohen.
\emph{Analyzing Neural Time Series Data: Theory and Practice}.
MIT Press, 2014.

\bibitem{oppenheim}
A.~V.~Oppenheim, J.~S.~Lim, B.~R.~Musicus, et~al.
Digital signal processing.
MIT, 1997.

\bibitem{welch1967}
P.~Welch.
The use of fast Fourier transform for the estimation of power spectra: A method based on time averaging over short, modified periodograms.
\emph{IEEE Trans.\ Audio Electroacoust.}, 15:70--73, 1967.

\bibitem{owens1988}
F.~J.~Owens and M.~S.~Murphy.
A short-time Fourier transform.
\emph{Signal Processing}, 14:3--10, 1988.

\bibitem{tallonbaudry1999}
C.~Tallon-Baudry and O.~Bertrand.
Oscillatory gamma activity in humans and its role in object representation.
\emph{Trends Cogn.\ Sci.}, 3:151--162, 1999.

\bibitem{boashash1992}
B.~Boashash.
Estimating and interpreting the instantaneous frequency of a signal.\ I.\ Fundamentals.
\emph{Proc.\ IEEE}, 80:520--538, 1992.

\bibitem{mitra1999}
P.~P.~Mitra and B.~Pesaran.
Analysis of dynamic brain imaging data.
\emph{Biophys.\ J.}, 76:691--708, 1999.

\bibitem{thomson1982}
D.~J.~Thomson.
Spectrum estimation and harmonic analysis.
\emph{Proc.\ IEEE}, 70:1055--1096, 1982.

\bibitem{harmony2013}
T.~Harmony.
The functional significance of delta oscillations in cognitive processing.
\emph{Front.\ Integr.\ Neurosci.}, 7:83, 2013.

\bibitem{buzsaki2004}
G.~Buzs\'aki and A.~Draguhn.
Neuronal oscillations in cortical networks.
\emph{Science}, 304:1926--1929, 2004.

\bibitem{klimesch1999}
W.~Klimesch.
EEG alpha and theta oscillations reflect cognitive and memory performance: a review and analysis.
\emph{Brain Res.\ Rev.}, 29:169--195, 1999.

\bibitem{cavanagh2014}
J.~F.~Cavanagh and M.~J.~Frank.
Frontal theta as a mechanism for cognitive control.
\emph{Trends Cogn.\ Sci.}, 18:414--421, 2014.

\bibitem{pfurtscheller1999}
G.~Pfurtscheller and F.~H.~Lopes~da~Silva.
Event-related EEG/MEG synchronization and desynchronization: basic principles.
\emph{Clin.\ Neurophysiol.}, 110:1842--1857, 1999.

\bibitem{klimesch2012}
W.~Klimesch.
$\alpha$-band oscillations, attention, and controlled access to stored information.
\emph{Trends Cogn.\ Sci.}, 16:606--617, 2012.

\bibitem{engel2010}
A.~K.~Engel and P.~Fries.
Beta-band oscillations---signalling the status quo?
\emph{Curr.\ Opin.\ Neurobiol.}, 20:156--165, 2010.

\bibitem{bastos2015}
A.~M.~Bastos and J.-M.~Schoffelen.
A tutorial review of functional connectivity analysis methods and their interpretational pitfalls.
\emph{Front.\ Syst.\ Neurosci.}, 9:175, 2015.

\bibitem{friston2011}
K.~J.~Friston.
Functional and Effective Connectivity: A Review.
\emph{Brain Connectivity}, 1:13--36, 2011.

\bibitem{nolte2004}
G.~Nolte, O.~Bai, L.~Wheaton, Z.~Mari, S.~Vorbach, and M.~Hallett.
Identifying true brain interaction from EEG data using the imaginary part of coherence.
\emph{Clin.\ Neurophysiol.}, 115:2292--2307, 2004.

\bibitem{lachaux1999}
J.-P.~Lachaux, E.~Rodriguez, J.~Martinerie, and F.~J.~Varela.
Measuring phase synchrony in brain signals.
\emph{Hum.\ Brain Mapp.}, 8:194--208, 1999.

\bibitem{stam2007}
C.~J.~Stam, G.~Nolte, and A.~Daffertshofer.
Phase lag index: assessment of functional connectivity from multi-channel EEG and MEG with diminished bias from common sources.
\emph{Hum.\ Brain Mapp.}, 28:1178--1193, 2007.

\bibitem{granger1969}
C.~W.~J.~Granger.
Investigating causal relations by econometric models and cross-spectral methods.
\emph{Econometrica}, 37:424, 1969.

\bibitem{kaminski1991}
M.~J.~Kami\'nski and K.~J.~Blinowska.
A new method of the description of the information flow in the brain structures.
\emph{Biol.\ Cybern.}, 65:203--210, 1991.

\bibitem{blinowska2011}
K.~J.~Blinowska.
Review of the methods of determination of directed connectivity from multichannel data.
\emph{Med.\ Biol.\ Eng.\ Comput.}, 49:521--529, 2011.

\bibitem{baillet2001}
S.~Baillet, J.~C.~Mosher, and R.~M.~Leahy.
Electromagnetic brain mapping.
\emph{IEEE Signal Process.\ Mag.}, 18:14--30, 2001.

\bibitem{medani2025editorial}
T.~Medani, S.~Pursiainen, M.-C.~Piastra, J.~Vorwerk, and R.~M.~Leahy.
Editorial: Forward and inverse solvers in multi-modal electric and magnetic brain imaging.
\emph{Front.\ Hum.\ Neurosci.}, 19:1629489, 2025.

\bibitem{kao2015}
J.~C.~Kao, P.~Nuyujukian, S.~I.~Ryu, M.~M.~Churchland, J.~P.~Cunningham, and K.~V.~Shenoy.
Single-trial dynamics of motor cortex and their applications to brain-machine interfaces.
\emph{Nat.\ Commun.}, 6:7759, 2015.

\bibitem{blankertz2011}
B.~Blankertz, S.~Lemm, M.~Treder, S.~Haufe, and K.-R.~M\"uller.
Single-trial analysis and classification of ERP components---a tutorial.
\emph{NeuroImage}, 56:814--825, 2011.

\bibitem{grootswagers2017}
T.~Grootswagers, S.~G.~Wardle, and T.~A.~Carlson.
Decoding dynamic brain patterns from evoked responses: A tutorial on multivariate pattern analysis applied to time series neuroimaging data.
\emph{J.\ Cogn.\ Neurosci.}, 29:677--697, 2017.

\bibitem{jeong2025decoding}
W.~Jeong, A.~Kommineni, K.~Avramidis, C.~McDaniel, D.~Berry, M.~Hughes, et~al.
Time-resolved {EEG} decoding reveals altered neural dynamics of affective semantic evaluation in depression and suicidality.
\emph{Commun.\ Biol.}, 2026.

\bibitem{jeong2023multivariate}
W.~Jeong, S.~Kim, J.~Park, and J.~Lee.
Multivariate EEG activity reflects the Bayesian integration and the integrated Galilean relative velocity of sensory motion during sensorimotor behavior.
\emph{Commun.\ Biol.}, 6:113, 2023.

\bibitem{kommineni2025}
A.~Kommineni, W.~Jeong, K.~Avramidis, C.~McDaniel, M.~Hughes, T.~McGee, et~al.
Neural responses to affective sentences reveal signatures of depression.
\emph{Transl.\ Psychiatry}, 2026.

\bibitem{kriegeskorte2008}
N.~Kriegeskorte.
Representational similarity analysis---connecting the branches of systems neuroscience.
\emph{Front.\ Syst.\ Neurosci.}, 2008.

\bibitem{cichy2014}
R.~M.~Cichy, D.~Pantazis, and A.~Oliva.
Resolving human object recognition in space and time.
\emph{Nat.\ Neurosci.}, 17:455--462, 2014.

\bibitem{king2014}
J.-R.~King and S.~Dehaene.
Characterizing the dynamics of mental representations: the temporal generalization method.
\emph{Trends Cogn.\ Sci.}, 18:203--210, 2014.

\bibitem{durbin2012}
J.~Durbin and S.~J.~Koopman.
\emph{Time Series Analysis by State Space Methods}.
Oxford University Press, 2nd edition, 2012.

\bibitem{cheung2010}
B.~L.~P.~Cheung, B.~A.~Riedner, G.~Tononi, and B.~D.~Van~Veen.
Estimation of cortical connectivity from EEG using state-space models.
\emph{IEEE Trans.\ Biomed.\ Eng.}, 57:2122--2134, 2010.

\bibitem{galka2010}
A.~Galka, K.~K.~F.~Wong, and T.~Ozaki.
Generalized state-space models for modeling nonstationary EEG time-series.
In \emph{Modeling Phase Transitions in the Brain}, pp.\ 27--52. Springer, 2010.

\bibitem{brouwer2009}
G.~J.~Brouwer and D.~J.~Heeger.
Decoding and Reconstructing Color from Responses in Human Visual Cortex.
\emph{J.\ Neurosci.}, 29:13992--14003, 2009.

\bibitem{brouwer2011}
G.~J.~Brouwer and D.~J.~Heeger.
Cross-orientation suppression in human visual cortex.
\emph{J.\ Neurophysiol.}, 106:2108--2119, 2011.

\bibitem{jeong2019}
W.~Jeong, S.~Kim, Y.-J.~Kim, and J.~Lee.
Motion-direction representation in multivariate electroencephalographic activity during smooth pursuit eye movements.
\emph{NeuroImage}, 202:116160, 2019.

\bibitem{foster2016}
J.~J.~Foster, D.~W.~Sutterer, J.~T.~Serences, E.~K.~Vogel, and E.~Awh.
The topography of alpha-band activity tracks the content of spatial working memory.
\emph{J.\ Neurophysiol.}, 115:168--177, 2016.

\bibitem{foster2017}
J.~J.~Foster, E.~M.~Bsales, R.~J.~Jaffe, and E.~Awh.
Alpha-Band Activity Reveals Spontaneous Representations of Spatial Position in Visual Working Memory.
\emph{Curr.\ Biol.}, 27:3216--3223.e6, 2017.

\bibitem{student1908}
Student.
The Probable Error of a Mean.
\emph{Biometrika}, 6:1, 1908.

\bibitem{fisher1992}
R.~A.~Fisher.
Statistical methods for research workers.
In \emph{Breakthroughs in Statistics}, pp.\ 66--70. Springer, 1992.

\bibitem{nichols2002}
T.~E.~Nichols and A.~P.~Holmes.
Nonparametric permutation tests for functional neuroimaging: a primer with examples.
\emph{Hum.\ Brain Mapp.}, 15:1--25, 2002.

\bibitem{maris2007}
E.~Maris and R.~Oostenveld.
Nonparametric statistical testing of EEG- and MEG-data.
\emph{J.\ Neurosci.\ Methods}, 164:177--190, 2007.

\bibitem{dunn1961}
O.~J.~Dunn.
Multiple comparisons among means.
\emph{J.\ Am.\ Stat.\ Assoc.}, 56:52--64, 1961.

\bibitem{benjamini1995}
Y.~Benjamini and Y.~Hochberg.
Controlling the false discovery rate: A practical and powerful approach to multiple testing.
\emph{J.\ R.\ Stat.\ Soc.\ B}, 57:289--300, 1995.

Classifiers and Beyond
\bibitem{devlin2019bert}
J.~Devlin, M.-W.~Chang, K.~Lee, and K.~Toutanova.
{BERT}: Pre-training of deep bidirectional transformers for language understanding.
In \emph{Proc.\ NAACL-HLT}, pp.\ 4171--4186, 2019.

\bibitem{baevski2020wav2vec}
A.~Baevski, Y.~Zhou, A.~Mohamed, and M.~Auli.
wav2vec 2.0: A framework for self-supervised learning of speech representations.
\emph{Adv.\ Neural Inf.\ Process.\ Syst.}, 33:12449--12460, 2020.

\bibitem{radford2021learning}
A.~Radford, J.~W.~Kim, C.~Hallacy, A.~Ramesh, G.~Goh, S.~Agarwal, et~al.
Learning transferable visual models from natural language supervision.
In \emph{Proc.\ ICML}, pp.\ 8748--8763. PMLR, 2021.

\bibitem{radford2023robust}
A.~Radford, J.~W.~Kim, T.~Xu, G.~Brockman, C.~McLeavey, and I.~Sutskever.
Robust speech recognition via large-scale weak supervision.
In \emph{Proc.\ ICML}, pp.\ 28492--28518. PMLR, 2023.

\bibitem{brown2020language}
T.~Brown, B.~Mann, N.~Ryder, M.~Subbiah, J.~D.~Kaplan, P.~Dhariwal, et~al.
Language models are few-shot learners.
\emph{Adv.\ Neural Inf.\ Process.\ Syst.}, 33:1877--1901, 2020.

\bibitem{jayaram2018moabb}
V.~Jayaram and A.~Barachant.
{MOABB}: trustworthy algorithm benchmarking for {BCIs}.
\emph{J.\ Neural Eng.}, 15(6):066011, 2018.

\bibitem{subasi2010eeg}
A.~Subasi and M.~I.~Gursoy.
{EEG} signal classification using {PCA}, {ICA}, {LDA} and support vector machines.
\emph{Expert Syst.\ Appl.}, 37(12):8659--8666, 2010.

\bibitem{bhattacharyya2010performance}
S.~Bhattacharyya, A.~Khasnobish, S.~Chatterjee, A.~Konar, and D.~N.~Tibarewala.
Performance analysis of {LDA}, {QDA} and {KNN} algorithms in left-right limb movement classification from {EEG} data.
In \emph{Proc.\ IEEE IC SMB}, pp.\ 126--131, 2010.

\bibitem{guler2007multiclass}
I.~Guler and E.~D.~Ubeyli.
Multiclass support vector machines for {EEG}-signals classification.
\emph{IEEE Trans.\ Inf.\ Technol.\ Biomed.}, 11(2):117--126, 2007.

\bibitem{murugavel2016hierarchical}
A.~S.~M.~Murugavel and S.~Ramakrishnan.
Hierarchical multi-class {SVM} with {ELM} kernel for epileptic {EEG} signal classification.
\emph{Med.\ Biol.\ Eng.\ Comput.}, 54(1):149--161, 2016.

\bibitem{Lawhern2016EEGNetAC}
V.~J.~Lawhern, A.~J.~Solon, N.~R.~Waytowich, S.~M.~Gordon, C.~P.~Hung, and B.~Lance.
{EEGNet}: a compact convolutional neural network for {EEG}-based brain--computer interfaces.
\emph{J.\ Neural Eng.}, 15(5):056013, 2018.

\bibitem{jing2023development}
J.~Jing, W.~Ge, S.~Hong, M.~B.~Fernandes, Z.~Lin, C.~Yang, et~al.
Development of expert-level classification of seizures and rhythmic and periodic patterns during {EEG} interpretation.
\emph{Neurology}, 100(17):e1750--e1762, 2023.

\bibitem{song2021transformer}
Y.~Song, X.~Jia, L.~Yang, and L.~Xie.
Transformer-based spatial-temporal feature learning for {EEG} decoding.
\emph{arXiv preprint arXiv:2106.11170}, 2021.

\bibitem{song2018eeg}
T.~Song, W.~Zheng, P.~Song, and Z.~Cui.
{EEG} emotion recognition using dynamical graph convolutional neural networks.
\emph{IEEE Trans.\ Affect.\ Comput.}, 11(3):532--541, 2018.

\bibitem{xiao2025dynamical}
Y.~Xiao, W.~Zheng, and G.~Zhao.
Dynamical Causal Graph Neural Network for {EEG} Emotion Recognition.
\emph{IEEE Trans.\ Affect.\ Comput.}, 2025.

\bibitem{obeid2016temple}
I.~Obeid and J.~Picone.
The {Temple University Hospital EEG} data corpus.
\emph{Front.\ Neurosci.}, 10:196, 2016.

\bibitem{kostas2021bendr}
D.~Kostas, S.~Aroca-Ouellette, and F.~Rudzicz.
{BENDR}: Using transformers and a contrastive self-supervised learning task to learn from massive amounts of {EEG} data.
\emph{Front.\ Hum.\ Neurosci.}, 15:653659, 2021.

\bibitem{yang2023biot}
C.~Yang, M.~Westover, and J.~Sun.
{BIOT}: Biosignal transformer for cross-data learning in the wild.
\emph{Adv.\ Neural Inf.\ Process.\ Syst.}, 36:78240--78260, 2023.

\bibitem{zhou2025csbrain}
Y.~Zhou, J.~Wu, Z.~Ren, Z.~Yao, W.~Lu, K.~Peng, et~al.
{CSBrain}: A cross-scale spatiotemporal brain foundation model for {EEG} decoding.
\emph{arXiv preprint arXiv:2506.23075}, 2025.

\bibitem{wang2024cbramod}
J.~Wang, S.~Zhao, Z.~Luo, Y.~Zhou, H.~Jiang, S.~Li, et~al.
{CBraMod}: A criss-cross brain foundation model for {EEG} decoding.
\emph{arXiv preprint arXiv:2412.07236}, 2024.

\bibitem{kommineni2024knowledge}
A.~Kommineni, K.~Avramidis, R.~Leahy, and S.~Narayanan.
Knowledge-guided {EEG} representation learning.
In \emph{Proc.\ IEEE EMBC}, pp.\ 1--6, 2024.

\bibitem{gu2021efficiently}
A.~Gu, K.~Goel, and C.~R\'e.
Efficiently modeling long sequences with structured state spaces.
\emph{arXiv preprint arXiv:2111.00396}, 2021.

\bibitem{avramidis2025neural}
K.~Avramidis, T.~Feng, W.~Jeong, J.~Lee, W.~Cui, R.~M.~Leahy, and S.~Narayanan.
Neural codecs as biosignal tokenizers.
\emph{arXiv preprint arXiv:2510.09095}, 2025.


\bibitem{cawley2010}
G.~Cawley and N.~L.~C.~Talbot.
On over-fitting in model selection and subsequent selection bias in performance evaluation.
\emph{J.\ Mach.\ Learn.\ Res.}, 11:2079--2107, Aug.\ 2010.

\bibitem{rommel2022}
C.~Rommel, J.~Paillard, T.~Moreau, and A.~Gramfort.
Data augmentation for learning predictive models on EEG: a systematic comparison.
\emph{J.\ Neural Eng.}, 19(6):066020, 2022.

\bibitem{jayaram2018}
V.~Jayaram and A.~Barachant.
MOABB: trustworthy algorithm benchmarking for BCIs.
\emph{J.\ Neural Eng.}, 15(6):066011, 2018

\end{thebibliography}
\end{document}